\documentclass{acm_proc_article-sp}

\usepackage{microtype}

\usepackage{listings}

\usepackage{subfig}

\usepackage{tikz}
\usetikzlibrary{matrix,positioning}

\definecolor{redi}{RGB}{255,38,0}
\definecolor{redii}{RGB}{200,50,30}
\definecolor{yellowi}{RGB}{255,251,0}
\definecolor{bluei}{RGB}{0,150,255}
\definecolor{blueii}{RGB}{135,247,210}
\definecolor{blueiii}{RGB}{91,205,250}
\definecolor{blueiv}{RGB}{115,244,253}
\definecolor{bluev}{RGB}{1,58,215}
\definecolor{orangei}{RGB}{240,143,50}
\definecolor{yellowii}{RGB}{222,247,100}
\definecolor{greeni}{RGB}{166,247,166}

\tikzset{ 
table/.style={
  matrix of nodes,
  row sep=-\pgflinewidth,
  column sep=-\pgflinewidth,
  nodes={rectangle,draw=black,text width=1.25ex,align=center},
  text depth=0.25ex,
  text height=1ex,
  nodes in empty cells
  },
texto/.style={font=\footnotesize\sffamily},
title/.style={font=\small\sffamily}
}

\newcommand\CellText[2]{%
  \node[texto,left=of mat#1,anchor=east]
  at (mat#1.west)
  {#2};
}

\newcommand\SlText[2]{%
  \node[texto,left=of mat#1,anchor=west,rotate=75]
  at ([xshift=3ex]mat#1.north)
  {#2};
}

\newcommand\RowTitle[2]{%
\node[title,left=6.3cm of mat#1,anchor=west]
  at (mat#1.north west)
  {#2};
}

\newcommand{\eg}{e.\,g.}
\newcommand{\ie}{i.\,e.}

\usepackage{setspace}

\lstdefinelanguage{lila}{
	basicstyle=\normalfont\ttfamily\small\singlespacing,
	keywordstyle={\bfseries},
	keywords={@aggregate,@split,@from,@to,@annotation,@enrich},
	breaklines=true,
	frame=lines,
	stepnumber=1,
	numbersep=5pt,
	tabsize=4,
	extendedchars=true, 
	captionpos=b,
	showspaces=false,
	showstringspaces=false
	xleftmargin=15pt
}

\begin{document}

\title{A Rule-based Language for Application Integration}
%
%
%
%
%

\numberofauthors{2} 
%
\author{
%
%
\alignauthor
Daniel Ritter\\
       \affaddr{SAP SE}\\
       \affaddr{Dietmar-Hopp-Allee 16}\\
       \affaddr{Walldorf, Germany}\\
       \email{daniel.ritter@sap.com}
\alignauthor
Jan Bro\ss\\
       \affaddr{Karlsruhe Institute of Technology}\\
       \affaddr{Kaiserstra\ss e 12}\\
       \affaddr{Karlsruhe, Germany}\\
       \email{jan.bross@student.kit.edu}
}

\maketitle
\begin{abstract}
Although message-based (business) application integration is based on orchestrated message flows, current modeling languages exclusively cover (parts of) the control flow, while under-specifying the data flow. Especially for more data-intensive integration scenarios, this fact adds to the inherent data processing weakness in conventional integration systems. 

We argue that with a more data-centric integration language and a relational logic based implementation of integration semantics, optimizations from the data management domain (\eg, data partitioning, parallelization) can be combined with common integration processing (\eg, scatter/gather, splitter/gather). With the \emph{Logic Integration Language} (LiLa) we re-define integration logic tailored for data-intensive processing and propose a novel approach to data-centric integration modeling, from which we derive the control-and data flow and apply them to a conventional integration system.
	
\end{abstract}




\section{Introduction} \label{sec:intro}
Conventional message-based integration systems show weaknesses when it comes to data-intensive (business) application integration (\eg, fast-growing business areas like online player position tracking in sports management, internet of things)--using integration patterns like message transformations and (partially) message routing \cite{DBLP:conf/dexa/RitterB14}. This is due to the facts that (\emph{P1)} most of the application data is stored in relational databases, which leads to many format conversions during the end-to-end processing, and the observation (\emph{P2}) that application tier programming languages (\eg, Java, C\#) are not (yet) table-centric. Promising solutions for that either ``push-down" integration logic to relational database systems \cite {DBLP:conf/zeus/Ritter14} or propose to re-define the integration patterns by relational (logic) programming \cite{DBLP:conf/dexa/RitterB14}. Both approaches promise efficient, data-intensive message processing, while allowing for optimizations with respect to common data- (\eg, data partitioning, parallelization) and message-based integration (\eg, scatter/gather \cite{Hohpe:2003:EIP:940308,DBLP:journals/is/RitterMR17}, splitter/gather), which have not been possible before.

At the same time, despite recent advances in control-and data flow modeling of workflow (\eg, \cite{DBLP:conf/birthday/AbiteboulV13}) and integration systems (\eg, \cite{DBLP:conf/ecmdafa/Ritter14,DBLP:journals/corr/Ritter14}), a data-centric modeling of integration scenarios is vacant. Currently, only the (mostly) control flow centric icon notation for the \emph{Enterprise Integration Patterns} (EIP) \cite{Hohpe:2003:EIP:940308,DBLP:journals/is/RitterMR17} can be considered a ``de-facto" modeling standard. Through the icons, the collected common integration patterns can be combined to describe and configure integration semantics on an abstract level.
\begin{figure*}[h!]	
	\centering
	\includegraphics[width=0.9\textwidth]{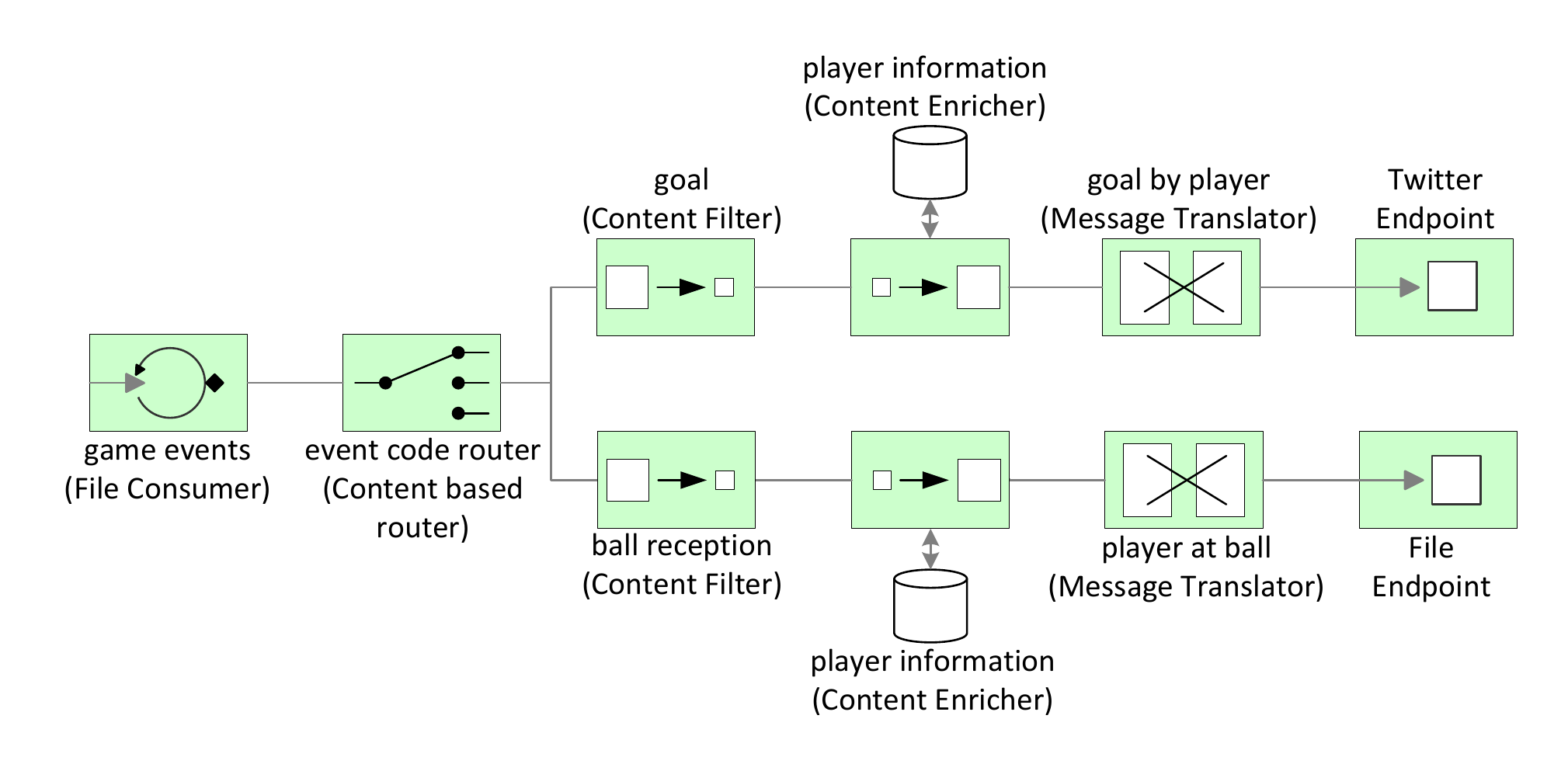}
	\caption{Excerpt from a soccer player event-message integration scenario (EIP icon notation \cite{Hohpe:2003:EIP:940308}).}
	\label{fig:twitter_bpmn}
\end{figure*} For instance, Figure \ref{fig:twitter_bpmn} shows a ``Soccer Player Event" integration scenario from sports management in the EIP icon notation. The player event data is gained through a \emph{Polling Consumer}, loading game events collected by sensors attached to the players and the playing field during a soccer match. Depending on the event code, a \emph{Content-based Router} pattern is used to route the messages to specific filter operations for ``Shots on goal" and ``Player at ball" through a \emph{Content Filter}, whereafter additional player information is merged into the resulting messages using a \emph{Content Enricher}. Then the messages are converted into the formats understood by their receivers using a \emph{Message Translator}. The ``Shots on goal" information is posted as twitter feed and ball possessions are stored to file. While the control flow is modeled, the message formats (\eg, ``Game events", ``Player information") and the actual data processing (\eg, routing and filter conditions, enricher and mapping programs) remain hidden on a second level configuration. In contrast, a more data-aware formalization should treat data as first-class citizen of an integration scenario. This (\emph{P3}) would give an integration expert the immediate control over the actual core aspect of integration, the data and its format, and (\emph{P4}) would take away the burden of explicitly modeling the system's control flow, while keeping best practices and optimizations in mind, which should rather be configured by the system itself. In this context there is a new trend to use Datalog-style rule-based languages to declaratively specify data-centric application development by Green et al \cite{DBLP:conf/datalog/GreenAK12} and Abiteboul et al \cite{DBLP:conf/sigmod/AbiteboulAMST13}, who applied logic programming (\ie, extended Datalog) to analytical and web application development. Similarly, we showed in \cite{DBLP:conf/dexa/RitterB14} the applicability and expressiveness of standard Datalog in the context of the EIPs \cite{Hohpe:2003:EIP:940308,DBLP:journals/is/RitterMR17}. 

To approach these observations (P1--P4), we propose a novel formalization tailored to data-intensive, message-based integration and a data-centric modeling approach, which we call \emph{Logic Integration Language} (LiLa). For that, we re-define core EIPs as part of a conventional integration system using Datalog. Datalog allows for data processing closer to its storage representation, and is sufficiently expressive for the evaluation of EIPs \cite{DBLP:conf/dexa/RitterB14}. Similarly, LiLa programs are based on standard Datalog$^+$, for which we carefully defined a small set of integration-specific extensions.
\begin{lstlisting}[language=lila,caption={Soccer Game Event Integration with LiLa.},label={lst:motivating-example}]
@from(file:gameEvents.json,json)
{gE(period,time,eventCode,pId).}

g(period,time,pId):-
   gE(period,time,"Goal",pId).
br(period,time,pId):-
   gE(period,time,"BallReception",pId).
   
gByP(period,time,firstN,lastN):-
   g(period,time,pId),pInfo(pId,firstN,lastN).
pAtB(period,time,firstN,lastN):-
   br(period,time,pId),pInfo(pId,firstN,lastN).

@enrich(playerInfo.json,json)
{pInfo(pId,firstN,lastN).}

@to(twitter:$config,json)
{gByP}
@to(file:playersAtBall.json)
{pAtB}
\end{lstlisting} For instance, Listing \ref{lst:motivating-example} shows the LiLa program of our motivating example. Notably, the data flow, formats and operations are represented as Datalog program with annotations. The file-based message adapter \texttt{@from} reads a stream of game events in the JSON format, canonically converts and projects the message body to Datalog facts of the form \texttt{gE}. Several Datalog rules represent operations on the data like filters (\ie, predicates \texttt{g}, \texttt{br}), enricher \texttt{@enrich}, loading and merging \texttt{pInfo} from \texttt{gByP} and \texttt{pByB}), before binding the IDB relations to receiver endpoints \texttt{@to} that only pass specified predicates and (canonically) convert them to the configured format (\eg, JSON).
From the LiLa programs we derive integration semantics and an efficient control flow using pattern detection.
To show the applicability of our approach to real-world integration scenarios and to conduct performance measurements, we synthesize LiLa programs to the open-source integration system Apache Camel \cite{apacheCamel13} that implements most of the integration semantics in form of EIPs. The results of the runtime analysis show that the usage of a more data-centric message processing is especially promising (a) for message transformations, while the routing efficiency remains similar to the conventional processing, and (b) from an end-to-end messaging point of view. Furthermore, the data-centric modeling with LiLa emphasizes the potential for optimizations and a novel modeling clarity compared to the existing control flow centric languages.

The remainder of the paper is organized along its contributions. Section \ref{sec:datapatterns} briefly describes the re-definition of EIPs as Datalog programs as foundation for the construction of LiLa in Section \ref{sec:lila}. The synthesis of LiLa programs to Apache Camel is explained in Section \ref{sec:synthesis} as basis for experimental evaluations discussed in 
Section \ref{sec:experimental}. Section \ref{sec:relatedWork} sets LiLa in context to related work and Section \ref{sec:conclusion} concludes the paper.


\section{Integration Patterns in a Nutshell} \label{sec:datapatterns}
The \emph{Enterprise Integration Patterns} (EIPs) \cite{Hohpe:2003:EIP:940308,DBLP:journals/is/RitterMR17} define operations on the header (\ie, payload's meta-data) and body (\ie, message payload) of a message, which are normally implemented in the integration system's host language (\eg, Java, C\#). Thereby the actual integration operation (\ie, the content developed by an integration expert like mapping programs and routing conditions) can be differentiated from the implementation of the runtime system that evaluates the content operations and processes their results. We re-define the content operations using Datalog and leave the runtime system (implementation) as is. The resulting set of operations and integration language additions, which we call \emph{Integration Logic Programming} (ILP) targets an enhancement of conventional integration systems for data-intensive processing, while preserving the general integration semantics like \emph{Quality of Service} (\eg, best effort, exactly once) and the \emph{Message Exchange Pattern} (\eg, one-way, two-way). In other words, the content part for the patterns is evaluated by a Datalog system, which is invoked by an integration system that processes the results.

\subsection{Canonical Data Model}
When connecting applications, various operations are executed on the transferred messages in a uniform way. The arriving messages are converted into an internal format understood by the pattern implementation, called \emph{Canonical Data Model} (CDM) \cite{Hohpe:2003:EIP:940308,DBLP:journals/is/RitterMR17}, before the messages are transformed to the target format. Hence, if a new application is added to the integration solution only conversions between the CDM and the application format have to be created. Consequently, for the re-definition of integration patterns, we define a CDM as \emph{Datalog Program}
, which consists of a set of facts, with an optional set of (supporting) rules as message body and a set of meta-facts that describes the actual data as header. The meta-facts encode the name of the fact's predicate and all parameter names within the relation as well as the position of each parameter. With that information, parameters can be accessed by name instead of position by Datalog rules (\eg, for selections, projections). 

\subsection{Relational Logic Integration Patterns}
Before re-defining the patterns integration semantics for routing and transformation patterns using Datalog, by $function_{ilp}$, let us recall some relevant, basic Datalog operations: \texttt{join}, \texttt{projection}, \texttt{union}, and \texttt{selection}. The join of two relations $r(x, y)$ and $s(y, z)$ on parameter $y$ is encoded as $j(x, y, z) \leftarrow r(x, y), s(y, z)$, which projects all three parameters to the resulting predicate $j$. More explicitly, a projection on parameter $x$ of relation $r(x, y)$ is encoded as $p(x) \leftarrow r(x, y)$. The union of $r(x, y)$ and $s(x, y)$ is $u(x, y) \leftarrow r(x, y).$ $u(x, y) \leftarrow s(x, y)$, which combines several relations to one. The selection $r(x, y)$ according to a built-in predicate $\phi(x,[const|z])$ is encoded as $s(x, y) \leftarrow r(x, y), \phi(x,[const|z])$, which only returns $s(x,y)$ records for which $\phi(x,[const|z])$ evaluates to \texttt{true} for a given constant value $const$ or a variable value $z$. Built-in predicates can be numerical, binary relations $\phi(x,const)$ like $<,>,<=,>=,=$ as well as string, binary relations like $equals, contains, starts with, ends with$, numerical expressions based on binary operators like $=,+,-,*,/$ (\eg, $x=p(y)+1$) and operations on relations like $y=max(p(x)),y=min(p(x))$, which would assign the maximal or the minimal value of a predicate $p$ to a parameter $y$.

\begin{figure}
	\begin{tikzpicture}[node distance =0pt and 0.5cm]
	\matrix[table] (mat11) 
	{
		|[fill=blue]| & & |[fill=blue]| & & \\
		|| & & & |[fill=blue]| & \\
		|| & & & & \\
		|[fill=bluei]| & |[fill=bluei]| & |[fill=blue]| & & \\
		|[fill=blue]| & & |[fill=blue]| & & \\
		|| & & & & |[fill=blue]| \\
	};
	
	
	\matrix[table,below=of mat11] (mat21) 
	{
		|[fill=bluei]| & |[fill=blue]| & & |[fill=blue]| & \\
		|[fill=blue]| & & |[fill=blue]| & |[fill=blue]| & \\
		|| & & & & |[fill=blue]| \\
	};
	
	
	
	\SlText{11-1-1}{built-in}
	\SlText{11-1-2}{join}
	\SlText{11-1-3}{selection}
	\SlText{11-1-4}{projection}
	\SlText{11-1-5}{union}
	
	
	\RowTitle{11}{Message Routing};
	\CellText{11-1-1}{Router, Filter: };
	\CellText{11-2-1}{Recipient List};
	\CellText{11-3-1}{Multicast, Join Router};
	\CellText{11-4-1}{Splitter};
	\CellText{11-5-1}{Correlation, Completion};
	\CellText{11-6-1}{Aggregation};
	
	\RowTitle{21}{Message Transformation};
	\CellText{21-1-1}{Message translator};
	\CellText{21-2-1}{Content filter};
	\CellText{21-3-1}{Content enricher};
	
	\end{tikzpicture}
	\caption{Message routing and transformation patterns mapped to Datalog. Most common Datalog operations for a single pattern are marked ``dark blue", less common ones ``light blue", and possible but uncommon ones ``white".} \label{fig:routetransform}
\end{figure}
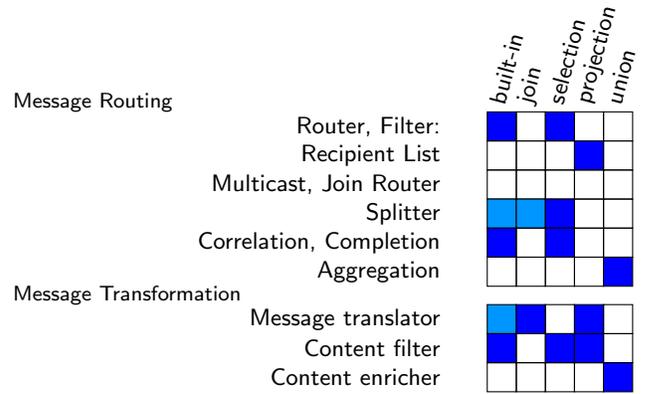

Although our approach allows each, single pattern definition to evaluate arbitrary Datalog rules, queries and built-in predicates, the Datalog to pattern mapping tries to identify and focus on the most relevant Datalog operations for a specific pattern. An overview of all discussed, re-defined routing functions and their mapping to Datalog constructs is shown in Figure \ref{fig:routetransform}.

\paragraph{Message Routing Patterns}
The routing patterns can be seen as control and data flow definitions of an integration channel pipeline. For that, they access the message to route it within the integration system and eventually to its receiver(s). They influence the channel and message cardinality as well as the content of the message. The most common routing pattern that determines the message's route based on its body is the \emph{Content-based Router}. The stateless router has a channel cardinality of $1$:$n$, where $n$ is the number of leaving channels, while one channel enters the router, and a message cardinality of $1$:$1$. The entering message constitutes the leaving message according to the evaluation of a \emph{routing condition}. This condition is a function $rc$, with $\{out_1, out_2, ..., out_n\} := rc(msg_{in}.body.x,conds)$, where $msg_{in}$ determines the entering message and $body.x$ is an arbitrary field $x$ of its structure. The function $rc$ evaluates to a list of Boolean output on a list of conditions $conds$ (\ie, Datalog rules) for each leaving channel. The output $\{out_1, out_2, ..., out_n\}$ is a list of Boolean values for each of the $n \in N$ leaving channels. However, only one channel must evaluate to \texttt{true}, all others to \texttt{false}. The Boolean output determines on which leaving channel the message is routed further (\ie, exactly one channel will route the message). Common integration systems implement a routing function that provides the entering message $msg_{in}$, represented by a Datalog program (\ie, mostly facts) and the $conds$ configurations as Datalog rules. Since standard Datalog rules cannot directly produce a Boolean result, there are at least two ways of re-defining $rc$: (a) by a supporting function in the integration system, or (b) by adding Boolean Datalog facts for each leaving channel that are joined with the evaluated conditions and exclusively returned by projection (not further discussed). 
An additional function $help\_rc$ for option (a), could be defined as $\{out_1, out_2, ..., out_n\} := help\_rc(list(list(fact)))$, fitting to the input of the routing function, where $list(list(fact))$ describes the resulting facts of the evaluation of $conds$ for each channel. The function $help\_rc$ emits \texttt{true}, if and only if $list(facts)\neq\emptyset$, and \texttt{false} otherwise. Now, the ILP routing condition is defined as $list(fact) := ilp_{rc}(msg_{in}.body.x,conds)$, while being evaluated for each channel condition, thus generating $list(list(fact))$. The $conds$ would then mainly be Datalog operations like selection or built-in predicates. For the message filter, which is a special case of the router that distinguishes only in its channel cardinality of $1$:$1$ and the resulting message cardinality of $1$:$[0|1]$, the $ilp_{rc}$ would have to be be evaluated once.

The stateless \emph{Multicast} and \emph{Recipient List} patterns route multiple messages to leaving channels, which gives them a message and channel cardinality of $1$:$n$. While the multicast routes messages statically to the leaving channels, the recipient list determines the receiving channels dynamically. The receiver determination function $rd$, with \[\{out_1, out_2, ..., out_n\} := rd(msg_{in}.[header.y|body.x]). ,\] computes $n \in N$ receiver channel configurations $\{out_1, out_2, \linebreak..., out_n\}$ by extracting their key values either from an arbitrary message header field $y$ or from the body $x$ field of the message. The integration system has to implement a receiver determination function that takes the list of key-strings $\{out_1, out_2, ..., out_n\}$ as input, for which it looks up receiver configurations $recv_{i}, recv_{i+1}, ..., recv_{i+m}$, where $i,m,n \in N$ and $m \leq n$, and passes copies of the entering message $\{msg_{out}', msg_{out}'', ..., msg_{out}^{m'}\}$. In terms of Datalog, $rd_{ilp}$ is a projection from values of the message body or header to a unary, output relation. For instance, the receiver configuration keys $recv_1$ and $recv_2$ have to be part of the message body like $body(x,'recv_1').\allowbreak body(x,'recv_2').$ and $rd_{ilp}$ would evaluate a Datalog rule similar to $config(y) \leftarrow body(x, y)$. For more dynamic receiver determinations, a dynamic routing pattern could be used. However, deviations from the original pattern defined in \cite{Hohpe:2003:EIP:940308,DBLP:journals/is/RitterMR17} would extend the expressiveness of the recipient list. Our ILP definition does not prevent from doing that. The multicast and join router pattern are statically configurable $1$:$n$ and $n$:$1$ channel patterns, which do not need a re-definition as ILP.

The antipodal \emph{Splitter} and \emph{Aggregator} patterns both have a channel cardinality of $1$:$1$ and create new leaving messages. Thereby the splitter breaks the entering message into multiple (smaller) messages (\ie, message cardinality of $1$:$n$) and the aggregator combines multiple entering messages to one leaving message (\ie, message cardinality of $n$:$1$). To be able to receive multiple messages from different channels, a \emph{Join Router} \cite{DBLP:journals/corr/Ritter14} pattern with a channel cardinality of $n$:$1$ and message cardinality of $1$:$1$ can be used as predecessor to the aggregator. Hereby, the stateless splitter uses a split condition $sc$, with $\{out1, out2,..., outn\} \&= sc(msg_{in}.body,conds)$, which accesses the entering message's body to determine a list of distinct body parts $\{out_1, out_2, ..., out_n\}$, based on a list of conditions $conds$, that are each inserted to a list of individual, newly created, leaving messages $\{msg_{out1}, msg_{out2}, ..., msg_{outn}\}$ with $n \in N$ by a splitter function. The header and attachments are copied from the entering to each leaving message. The re-definition $sc_{ilp}$ of split condition $sc$ evaluates a set of Datalog rules as $conds$, which mostly use Datalog selection, and sometimes built-in and join constructs (the latter two are marked ``light blue"). Each part of the body $out_i$ is a set of facts that is passed to a split function, which wraps each set into a single message.

The stateful aggregator defines a correlation condition, completion condition and an aggregation strategy. The correlation condition $crc$, with $coll_i := crc(msg_{in}.[header.y|body.x], \allowbreak conds)$, determines the aggregate collection $coll_i$, with $i \in N$, based on a set of conditions $conds$ to which the message is stored. The completion condition $cpc$, with $cpout := cpc(msg_{in}.[header.y|body.x])$, evaluates to a Boolean output $cpout$ based on header or body field information (similar to the message filter). If $cpout==true$, then the aggregation strategy $as$, with $aggout := as({msg_{in1}, msg_{in2}, ..., msg_{inn}})$, is called by an implementation of the messaging system and executed, else the current message is added to the collection $coll_i$. The $as$ evaluates the correlated entering message collection $coll_i$ and emits a new leaving message $msg_{out}$. For that, the messaging system has to implement an aggregation function that takes $aggout$ (\ie, the output of $as$) as input. These three functions are re-defined as $crc_{ilp}$, $cpc_{ilp}$ such that the $conds$ are rules mainly with selection and built-in Datalog constructs. The $cpc_{ilp}$ makes use of the defined $help\_rc$ function to map its evaluation result (\ie, list of facts or empty) to the Boolean value $cpout$. The aggregation strategy $as$ is re-defined as $as_{ilp}$, which mainly uses Datalog union to combine lists of facts from different messages. The message format remains the same. To transform the aggregates' formats, a message translator should be used to keep the patterns modular. However, the combination of the aggregation strategy with translation capabilities could lead to runtime optimizations. An overview of all discussed, re-defined routing functions and their mapping to Datalog constructs is shown in Figure \ref{fig:routetransform}.

\paragraph{Message Transformation Patterns}
The transformation patterns exclusively target the content of the messages in terms of format conversations and modifications.

The stateless \emph{Message Translator} changes the structure or format of the entering message without generating a new one (\ie, channel, message cardinality $1$:$1$). For that, the translator computes the transformed structure by evaluating a mapping program $mt$, with $msg_{out}.body := mt(msg_{in}.body)$. Thereby the field content can be altered.

The related \emph{Content Filter} and \emph{Content Enricher} patterns can be subsumed by the general \emph{Content Modifier} pattern and share the same characteristics as the translator pattern. The filter evaluates a filter function $mt$, which only filters out parts of the message structure, \eg, fields or values, and the enricher adds new fields or values as $data$ to the existing content structure using an enricher program $ep$, with $msg_{out}.body := ep(msg_{in}.body,data)$.

The re-definition of the transformation function $mt_{ilp}$ for the message translator mainly uses Datalog join and projection (plus built-ins for numerical calculations and string operations, thus marked ``light blue") and Datalog selection, projection and built-in (mainly numerical expressions and character operations) for the content filter. While projections allow for rather static, structural filtering, the built-in and selection operators can be used to filter more dynamically based on the content. The resulting Datalog programs are passed as $msg_{out}.body$. In addition, the re-defined enricher program $ep_{ilp}$ mainly uses Datalog union operations to add additional $data$ to the message as Datalog programs. Figure \ref{fig:routetransform} summarizes the discussed message transformation functions.

\paragraph{Pattern Composition} The defined patterns can be composed to more complex integration programs (\ie, integration scenarios or pipelines). From the many combinations of patterns, we briefly discuss two important structural patterns that are frequently used in integration scenarios: (1) scatter/gather and (2) splitter/gather \cite{Hohpe:2003:EIP:940308,DBLP:journals/is/RitterMR17}. Both patterns are supported by the patterns re-defined as ILPs.

The scatter/gather pattern (with a $1$:$n$:$1$ channel cardinality) is a multicast or recipient list that copies messages to several, statically or dynamically determined pipeline configurations, which each evaluate a sequence of patterns on the messages in parallel. Through a join router and an aggregator pattern, the messages are structurally and content-wise joined.

The splitter/gather pattern (with a $1$:$n$:$1$ message cardinality) splits one message into multiple parts, which can be processed in parallel by a sequence of patterns. In contrast to the scatter/gather the pattern sequence is the same for each instance. A subsequently configured aggregator combines the messages to one.

%

\section{Logic Integration Language} \label{sec:lila}
In the context of data-intensive message-processing, the current control flow-centric integration languages do not allow to design the data flow. Through the re-definition of the integration patterns with Datalog as ILPs, a foundation for a data-centric definition of integration scenarios is provided. Hence the language design of the subsequently defined \emph{Logic Integration Language} (LiLa) is based on Datalog, which specify programs that carefully extend standard Datalog$^+$ by integration semantics using annotations for message endpoints and complex routing patterns. As shown in Listing \ref{lst:annotation-format}, an annotation consists of an head with name preceded by ``@" and zero or more parameters enclosed in brackets, as well as a body enclosed in curly brackets.
\begin{lstlisting}[caption={Format of an annotation in \emph{LiLa}},language=lila,label={lst:annotation-format},mathescape]
@<annotationName>(<parameter>$^+$)
{ <Annotation Body> }
\end{lstlisting}

\subsection{Logic Integration Language Programs}
A LiLa program defines dependencies between Datalog facts, rules and annotations similar to the dependency graph of a Datalog program \cite{DBLP:books/cs/Ullman88}. Let us recall that the cyclic dependency graph $DG_D$ of a (recursive) Datalog program is defined as $DG_D := (V_D,E_D)$, where the nodes $V_D$ of the graph are IDB predicates, and the edges $E_D$ are defined from a node $n_1 \in N$ (predicate 1) to a node $n_2 \in N$ (predicate 2), if and only if, there is a rule with predicate 1 in the head and predicate 2 in the body.
\begin{figure}[h!]	
	\centering
	\includegraphics[width=0.5\textwidth]{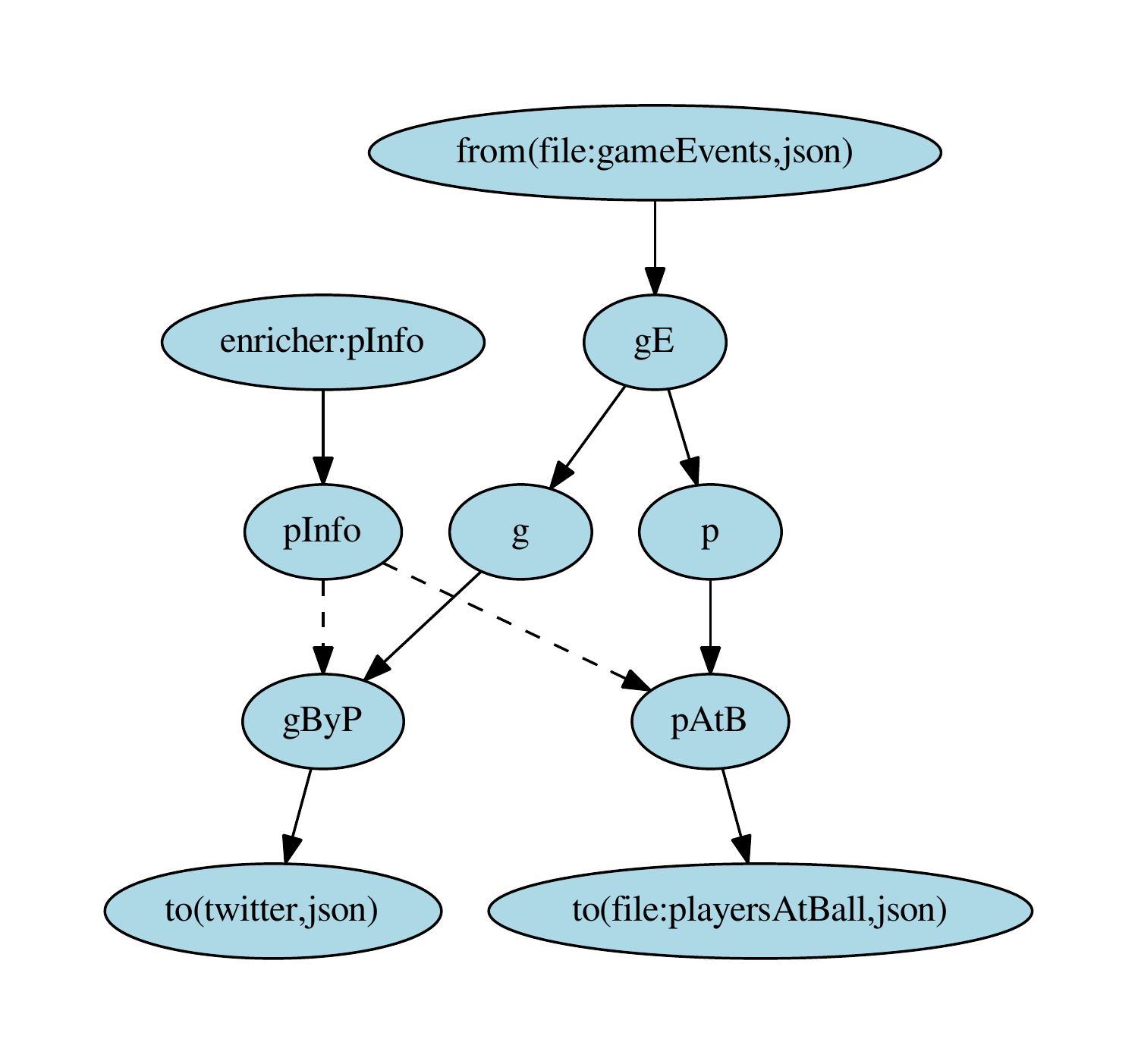}
	\caption{Dependency graph of the LiLa program from the motivating example.}
	\label{fig:simple-lila-dep-graph}
\end{figure}

Analogously, the directed, acyclic LiLa dependency graph $LDG$ is defined as $LDG := (V_p, E_p)$, where $V_p$ are collections of IDB predicates, which we call processors. An edge $E_p$ from processor $p_1 \in V_p$ to $p_2 \in V_p$ exists, if there is a rule with predicate 1 from $p_1$ in the head and predicate 2 from $p_2$ in the body. Hence the LDG contains processors with embedded cyclic rule dependency graphs, which do not lead to cycles in the $LDG$. In contrast to the $DG_D$, annotations are added to the $LDG$ as nodes. If an annotation uses a predicate an edge from that predicate is drawn to the node of the annotation (\ie, annotation depends on that predicate). If another annotation or rule uses the predicates produced by an annotation an edge from the annotation to the node representing the annotation or rule, which uses the data produced by the annotation, is drawn. Figure \ref{fig:simple-lila-dep-graph} shows the $LDG$ for the LiLa program depicted in Listing \ref{lst:motivating-example}. The message endpoint nodes are labeled with their consumer/producer URI with the predicate name of the rule for content filters.



\subsection{Endpoint-specific Extensions}
To connect the message sender, the \emph{Fact Source}, with the message receiver, the \emph{Routing Goal}, LiLa extends Datalog by \texttt{@from}, \texttt{@to} annotation statements similar to the open source integration system Apache Camel \cite{apacheCamel13}. Nodes of $LDG$ with no incoming edges are either EDB predicates or fact sources. Nodes with no outgoing edges are (mostly) routing goals. The only counter example are obsolete/unused processing steps, which can be deleted.

Representing the sender-facing fact source specifies the sender's transport and message protocol. Listing \ref{lst:fact-source-def} defines the fact source, which consists of a \texttt{location}, configuration URI that can be directly interpreted by an integration system and defines the location of the facts, and \texttt{format}, the message format of the data source (\eg, JSON, CSV, XML). The annotation body specifies the format's relations in form of Datalog facts. The message format is canonically converted to Datalog programs according to the ILP-CDM.

\begin{lstlisting}[caption={Definition of a fact source in \textsc{LiLa}},language=lila,label={lst:fact-source-def},mathescape]
@from(<location>,<format>)
{ <relationName(<parameter>$^+$)>.$^+$ }
\end{lstlisting}
\vspace{-0.1cm}

Similarily, the routing goal definitions specify the receiver-facing transport and message protocols (cf. Listing \ref{lst:routing-goal-def}). Hereby, the ILP-CDM is canonically converted to the message format understood by the receiver.

\begin{lstlisting}[caption={Definition of a routing goal in \textsc{LiLa}},language=LiLa,label={lst:routing-goal-def},mathescape]
@to(<producerURI>,<format>)
{ <relationName>[<linebreak><relationName>]$^*$ }
\end{lstlisting}

\subsection{Inherent Integration Patterns}
The Datalog facts provided by the fact source can be directly evaluated by Datalog rules. The LiLa dependency graph is used to automaticlly identify message transformation and basic routing patterns.

\paragraph{Message Transformation Patterns} 
Further patterns that can be derived from the $LDG$ are message transformation patterns like \emph{Content Filter}, \emph{Message Translator}, and the local \emph{Content Enricher}.

The content filter and message translator patterns are used to filter parts of a message as well as to translate the message's structure. Both are inherently declared in LiLa by using Datalog rules, which are collected in processors of the $LDG$. Each set of rules producing the same predicate corresponds to a filter or translator in the integration middleware. For instance, the LiLa program for the motivating example produces two content filters: one for the relation \texttt{gByP} and another one for the relation \texttt{pByB}. The routing between multiple content filters is decided based on the dependency graph of the LiLa program. If a node has a single outgoing edge, the incoming data is directly routed to the processor corresponding to the subsequent node. If a node has multiple incoming edges a join router pattern is present, which is detected and transformed as described in Section \ref{sec:synthesis}. The same is the case for a node having multiple outgoing edges, which corresponds to a multicast pattern.

For the local content enricher, LiLa allows to specify facts in a LiLa program. The facts are treated as processor (\ie, node in $LDG$) and are automatically placed into the message after a relation with this name is produced.

\paragraph{Message Routing Patterns}
In addition to the message transformation patterns, some simple routing patterns can be derived from the dependency graph like \emph{Multicast}, \emph{Message Filter}, \emph{Content-based Router} and \emph{Join Router}.

The multicast pattern can be used as part of the common map/reduce-style message processing. The multicast is derived by analyzing the dependency graph for independent rules to which copies of the message are provided. One potential side-effect is the detection of (too) many multicast configuration, when a routing goal requests multiple intermediary results of a single route, which we mitigate by an optimization that keeps these results (not shown).

The message filter, removes messages according to a filter condition (cf. Section \ref{sec:datapatterns}). For the filter, LiLa does not define a special construct The filtering of a message can be achieved by performing a content filtering, which leads to an empty message. Empty messages are discarded before sending the message for further processing to a routing goal. This behavior can be used to describe a content based-router, which distinguishes from the filter by its message cardinality of $1$:$n$. However in LiLa, we use the router with a channel cardinality of $1$:$n$ (\ie, multicast) with message filters on each leaving message channel according to \cite{Hohpe:2003:EIP:940308}.

A structurally, channel combining pattern is the join router. The join router has a channel cardinality of $n$:$1$, however, does only combine channels, but not messages. For that an aggregator is used that is defined subsequently.

\subsection{Routing-specific Extensions}
The more complex routing patterns \emph{Aggregator}, \emph{Splitter} and remote \emph{Content Enricher} can neither be described by standard Datalog nor inherently detected in the dependency graph. Hence we define special annotations for these patterns.

For the aggregator, Listing \ref{lst:aggregator-def} shows the \texttt{@aggregate} annotation with pre-defined aggregation strategies like \texttt{union} and an either time (\eg, \texttt{completionTime}=3) or number-of-messages based \texttt{completion condition} (\eg, \texttt{completionSize=5}). The annotation body consists of several Datalog queries. The message correlation is based on the query evaluation, where \texttt{true} means that the evaluation result is not an empty set of facts and \texttt{false} otherwise. As the aggregator does not produce facts with a new relation name, but combines multiple messages keeping their relations, it is challenging how to reference to the aggregated relations in a LiLa program as their name does not change (\ie, message producing). This leads to problems, when building the dependency graph, \ie, it is undecidable whether a rule uses the relation prior or after aggregation. As we do not want the user to specify explicitly, whether she means the relation prior or after aggregation in every rule using a predicate used in an aggregator, we suffix all predicates after an aggregation step with \texttt{-aggregate} by default. In combination with a join router, messages from several entering channels can be combined.
\begin{lstlisting}[caption={Definition of an aggregator in \textsc{LiLa}},language=lila,label={lst:aggregator-def},mathescape]
@aggregate(<aggregationStrategy>,<completionCondition>)
{ <?-<relationName>(<parameter>$^+$).>$^+$ }
\end{lstlisting}

LiLa specifies the splitter as in Listing \ref{lst:splitter-def} with a new \texttt{@split} annotation, which does not have any parameters in the annotation head. Datalog queries are used in the annotation body as splitter expressions as briefly described in Section \ref{sec:datapatterns}. The queries are evaluated on the exchange and each evaluation result is passed for further processing as a single message. Similar to the aggregator, all newly generated relations leaving a splitter are suffixed with \texttt{-split} by default in order to not explicitly having to specify, whether the relation prior or after splitting is meant.
\begin{lstlisting}[caption={Definition of a splitter in \textsc{LiLa}},language=lila,label={lst:splitter-def},mathescape]
@split()
{ <?-<relationName>(<parameter>$^+$).>$^+$ }
\end{lstlisting}

The remote content enricher can be seen as a special message endpoint. For instance for an enricher including data from a file, the \texttt{filename} and \texttt{format} have to be specified as shown in Listing \ref{lst:enrich-from-file}. Similar to the fact source, a set of relations has to be specified. Again, a canonical conversion from the specified file format to the ILP-CDM is conducted according to the definitions in Section \ref{sec:datapatterns}. If the relations to enrich via this construct are already generated by another construct or Datalog rule, they are enriched after this construct by adding the additional facts to the message. If there is no construct or Datalog rule producing the relations specified in the annotation body, the relations are enriched directly before their usage. The enricher construct is especially useful when a single message shall be combined with additional information.

\begin{lstlisting}[caption={Definition of a remote content enricher in \textsc{LiLa}},language=lila,label={lst:enrich-from-file},mathescape]
@enrich(<filename>,<format>)
{ <relationName(<parameter>$^+$).>$^+$ }
\end{lstlisting}

\section{Synthesis of Logic Integration Language Programs}\label{sec:synthesis}
The defined LiLa constructs can be combined to complex representations of integration programs, which can be executed by integration systems. For that we have chosen the lightweight, open-source integration system Apache Camel \cite{apacheCamel13}, since it implements all discussed integration semantics. To guarantee data-intensive processing, LiLa programs are not synthesized to Apache Camel constructs directly, but to the ILP integration pattern re-definitions that are plugged to the respective system implementations (cf. Section \ref{sec:datapatterns}). The equivalent to message channels in Apache Camel are Camel Routes.
\begin{figure}[h!]
	\centering
	\includegraphics[width=0.5\textwidth]{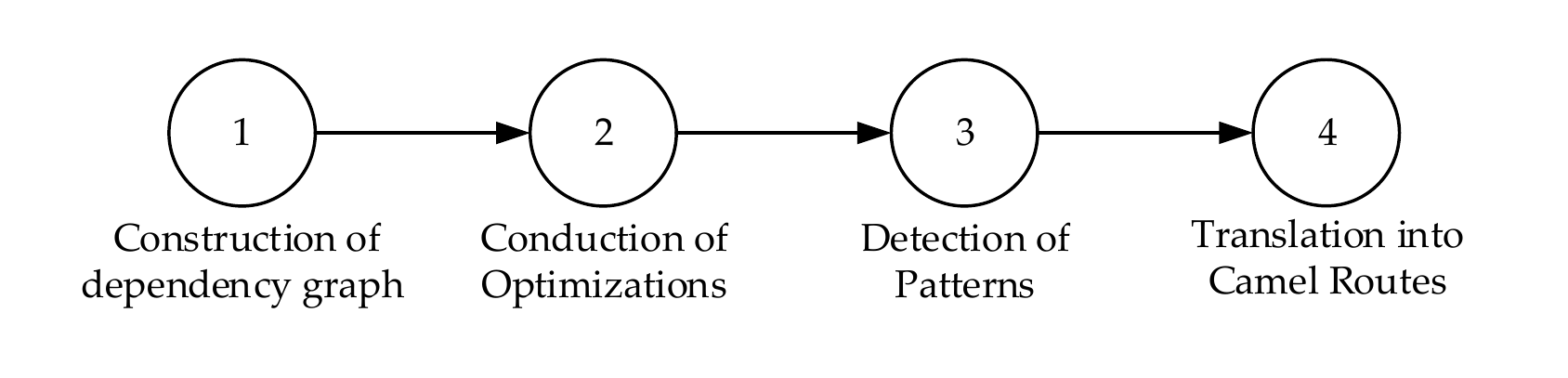}
	\caption[LiLa Compiler Pipeline]{LiLa Compiler Pipeline}
	\label{fig:compiler-pipeline}
\end{figure} 

\subsection{Message Channel/Route Graph}
The definition of a platform-independent message channel representation, called \emph{Route Graph} (RG), enables a graph transformation $t: LDG \rightarrow RG$ and an efficient code generation for different runtime systems. The transformation is a two-step process: In the first step a condition is evaluated on each edge or node respectively. If the condition evaluates to true, further processing on this node/edge is performed. The second step is the execution of the actual transformation.

The route graph $RG$ is defined as $RG$ := $(V_R, E_R)$, where the nodes $V_R$ are runtime components of an integration system (representing an ILP-EIP), and the edges $E_R$ are communication channels from one node $n_1 \in V_R$ to another node $n_2 \in V_R$, or itself $n_1$. In most integration systems acyclic route graphs are possible, however, not considered in this work. The nodes in $V_R$ can be partitioned to different routes, while edges in $E_R$ from one route to another have to be of type \texttt{to} for the source node and of type \texttt{from} for the target node. 

\begin{figure}[h!]
	\centering
	\includegraphics[width=0.5\textwidth]{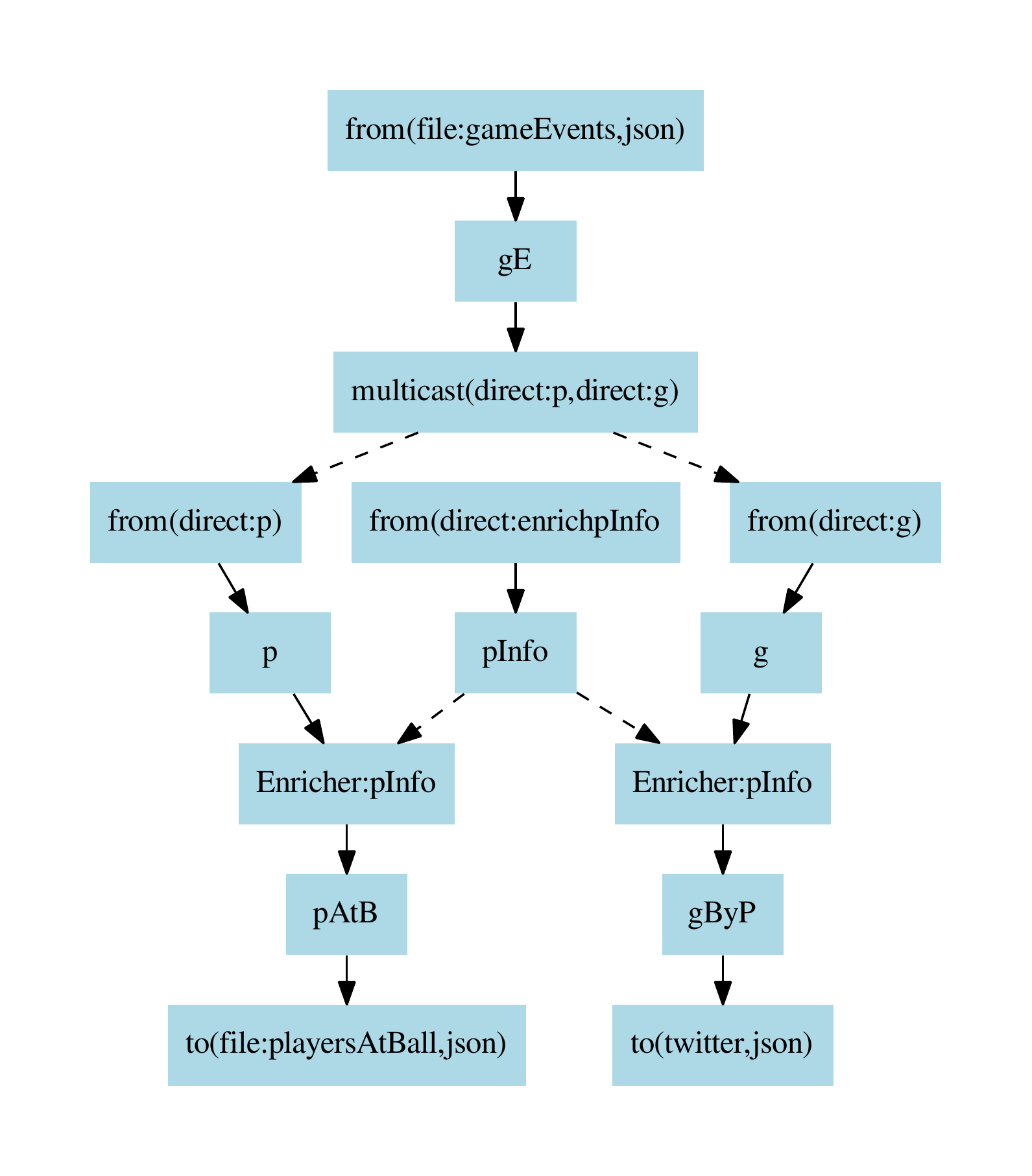}
	\caption[Route graph for the LiLa program of the motivating example.]{Route graph for the LiLa program of the motivating example in Figure \ref{fig:simple-lila-dep-graph}.}
	\label{fig:simple-lila-route-graph}
\end{figure}

For instance, Figure \ref{fig:simple-lila-route-graph} shows the route graph the $LDG$ of our motivating example (cf. Figure \ref{fig:simple-lila-dep-graph}). The message flow between separately generated routes (dashed-lines) indicate a \texttt{to}/\texttt{from} construct. Consequently, the LiLa program from Listing \ref{lst:motivating-example} results to four distinct routes, \eg, with a multicast \texttt{multicast(direct:p,direct:g)} and a file-enricher \texttt{from(direct:pEnrichInfo)} identified through pattern detection, which is subsequently discussed.

\subsection{Pattern Detection and Transformation}
The more complex, structural join router, multicast and remote enricher patterns are automatically derived from the $LDG$ through a rule-based pattern detection approach. With these building blocks, common optimizations in integration systems, \eg, the map/reduce-like scatter/gather pattern \cite{Hohpe:2003:EIP:940308,DBLP:journals/is/RitterMR17}, which is a combination of the multicast, join router and aggregator patterns, can be synthesized. The rule-based detection and transformation approach defines a matching function $[true|false] := mf_{LDG, mc}$ on $LDG$, with matching condition $mc$ and a transformation $t_G$, with $t_G: LDG \rightarrow RG$. The matching function denotes a node and edge graph traversal on the $LDG$ that evaluates to \texttt{true} if the condition holds, \texttt{false} otherwise. The transformation $t_G$ is executed only if the condition holds.

\begin{figure*}[h]
	\centering
	\subfloat[LiLa Graph]{\label{fig:joint-router-detection}\includegraphics[width=0.3\textwidth]{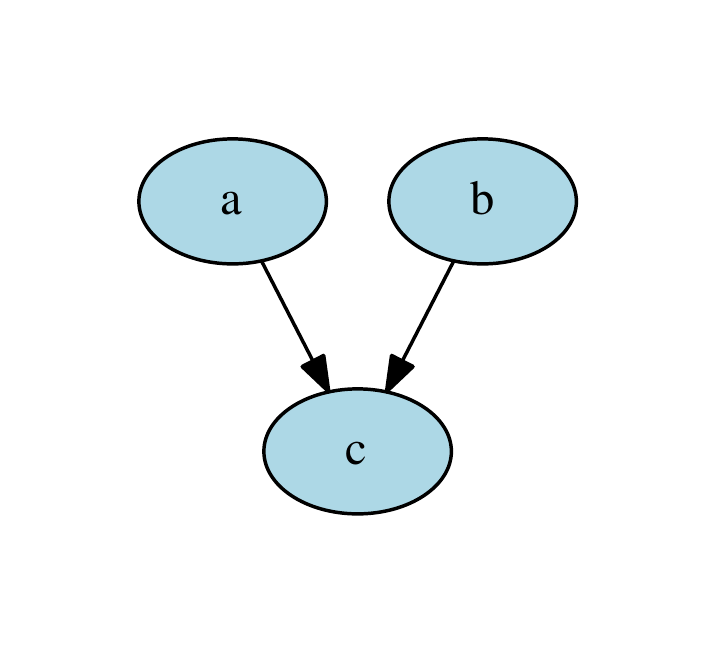}} 
	\subfloat[Route Graph]{\label{fig:joint-router-detected}\includegraphics[width=0.4\textwidth]{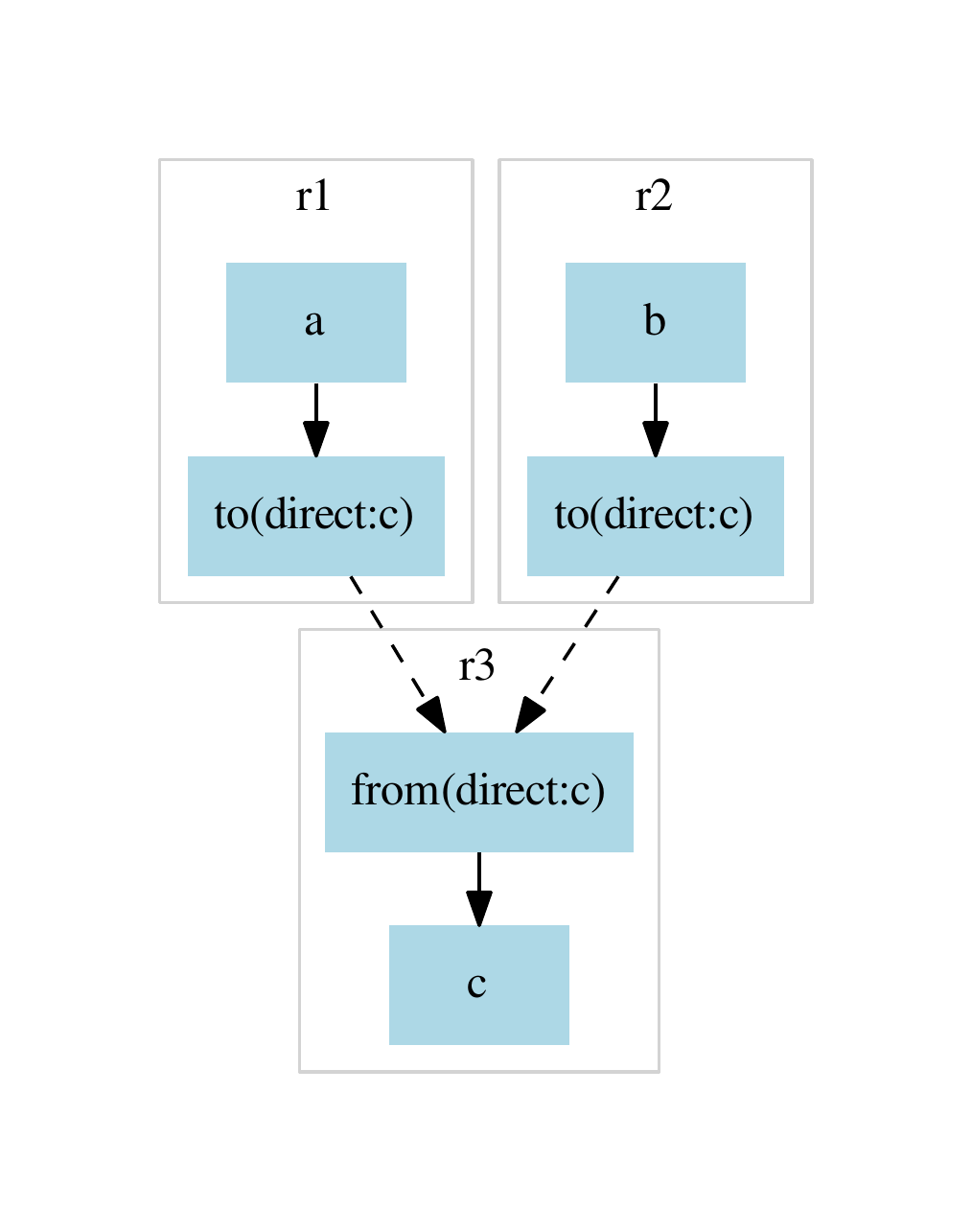}}
	\caption{Detection of Join Router Pattern.}
\end{figure*}

\paragraph{Join Router} The router is a $m$:$1$ message channel join pattern, which usually has to be combined with an aggregator to join messages. The match condition is defined as $mc_{JR} := deg^-(n_i) > 1$, where $deg^-(n_i)$ determines the number of entering message channels on a specific node $n_i \in V_P$, with $i \in N$. Hence, only in case of multiple entering edges, the graph transformation $t_{jr}$ is executed. The transformations $t_{jr_{1-3}}$ change the $RG$: $t_{jr_1}: n_i \rightarrow n_{fd} \oplus n_i$, for all matching nodes $n_i$, a \texttt{from-direct} node $n_{fd}$ is added, denoted by $\oplus$. Additionally, all nodes $n_j$ with direct, outgoing edges to the matching node get an additional \texttt{to-direct} node $n_{td}$: $t_{jr_2}: n_j \rightarrow n_j \oplus n_{td}$. Then all original edges $e_m$ have to be removed: $t_{jr_3}: E_P \rightarrow E_P\setminus e_m$. Figure \ref{fig:joint-router-detection} shows the $LDG$ and Figure \ref{fig:joint-router-detected} the corresponding $RG$ after the transformation.

\begin{figure*}[h]
	\centering
	\subfloat[LiLa Graph\label{fig:multicast-detection}]{\includegraphics[width=0.3\textwidth]{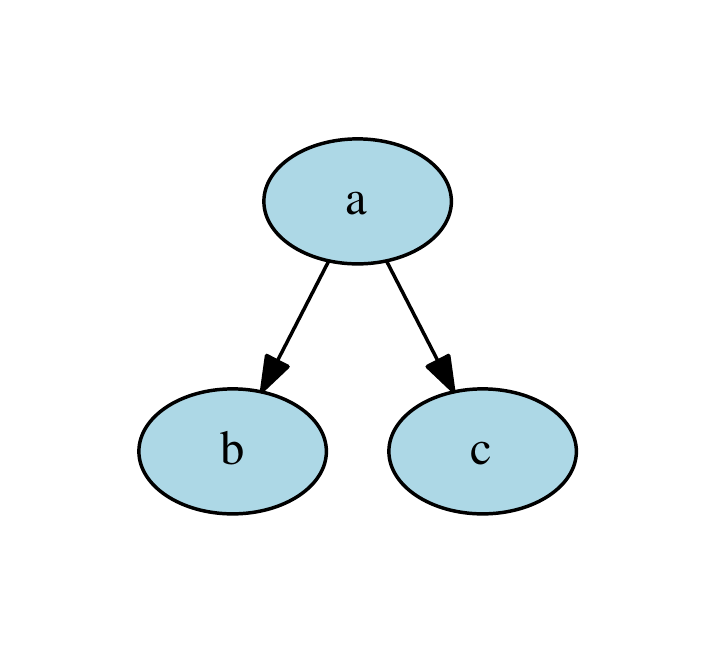}} 
	\subfloat[Route Graph\label{fig:multicast-detected}]{\includegraphics[width=0.4\textwidth]{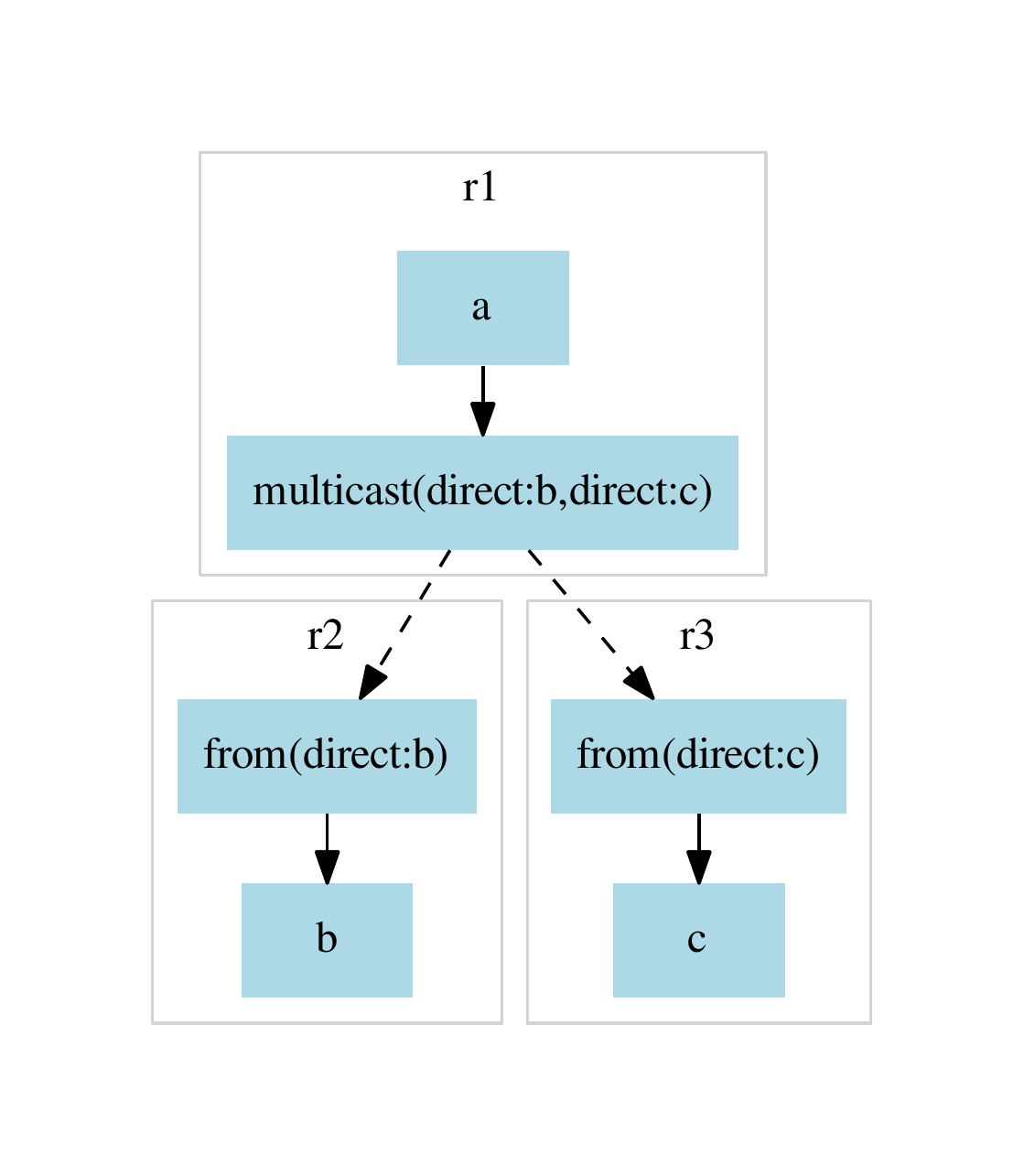}}
	\caption{Detection of Multicast Pattern.}
	\label{fig:multicast-graph-rule}
\end{figure*}

\paragraph{Multicast} The multicast has a channel cardinality of $1$:$n$. The match condition is defined as $mc_{Mu} := deg^+(n_i) > 1$, where $deg^+(n_i)$ determines the number of leaving message channels on a specific node $n_i \in V_P$, with $i \in N$. Hence, only in case of multiple leaving edges, the graph transformation $t_{jr}$ is executed. The transformations $t_{mu_{1--3}}$ change the $RG$: $t_{mu_1}: n_i \rightarrow n_i \oplus n_{multic\{n_j\}}$, for all matching nodes $n_i$, a \texttt{multicast} node $n_{multic}$ is added, which references all previous neighboring nodes $n_j$ via leaving edges. Then a \texttt{from-direct} node $n_{fd}$ is added to all neighboring nodes $n_j$ through transformation $t_{mu_2}: n_j \rightarrow n_{fd} \oplus n_j$. Additionally, all original edges $e_m$ have to be removed: $t_{mu_3}: E_P \rightarrow E_P\setminus e_m$. Figure \ref{fig:multicast-detection} shows the $LDG$ and Figure \ref{fig:multicast-detected} the corresponding $RG$ after the transformation.

\paragraph{Remote Enricher} An enricher potentially merges several predicate relations to the main route as additional data. Therefore it has to get a route on its own that (periodically) gathers the respective messages. The intermediate transformation $t_{re}$ is defined as $t_{re_1}: n_{i} \oplus \{n_{j}\} \rightarrow n_{fd} \oplus n_{file} \oplus \{n_{j}\}$, with $n_i, n_j \in V_P$, which takes all matching enricher nodes $n_i$ and the list of connected nodes $\{n_{j}\}$ and translates them to a \texttt{from-direct} node $n_{fd}$ that is followed by a file relation $n_{file}$, referencing the connected nodes $\{n_{j}\}$. Additionally, all original edges $e_m$ from the enricher $n_i$ to the list of connected nodes $\{n_{j}\}$ have to be removed: $t_{er_2}: E_P \rightarrow E_P\setminus e_m$. The match condition for the remote enricher is the node type $type(n_i)$, determined through the \texttt{@enrich} annotation: $mc_{RE} := type(n_i) == '@enrich'$. After the intermediate translation, all produced relations (nodes) that are linked to nodes in the main tree create a join router (cf. transformations $t_{jr_{1-3}}$) with a built-in aggregator that merges the facts, \eg, via union operation. Figure \ref{fig:multicast-detection} denotes the $LDG$ of an example remote enricher pattern that is transformed to its corresponding $RG$, shown in Figure \ref{fig:multicast-detected}. In order to find the complete path of nodes to extract the leaving edges have to be followed starting at the enricher node until a node that has multiple incoming nodes. Before the node with multiple incoming nodes, a \texttt{to-direct} node is inserted through $t_{jr_2}$ (dashed line). The URI of the call to enricher node is set to the URI of the consumer, which has to be added directly before the enricher node.

\begin{figure*}[h]
	\centering
	\subfloat[LiLa Graph\label{fig:enricher-detection}]{\includegraphics[width=0.5\textwidth]{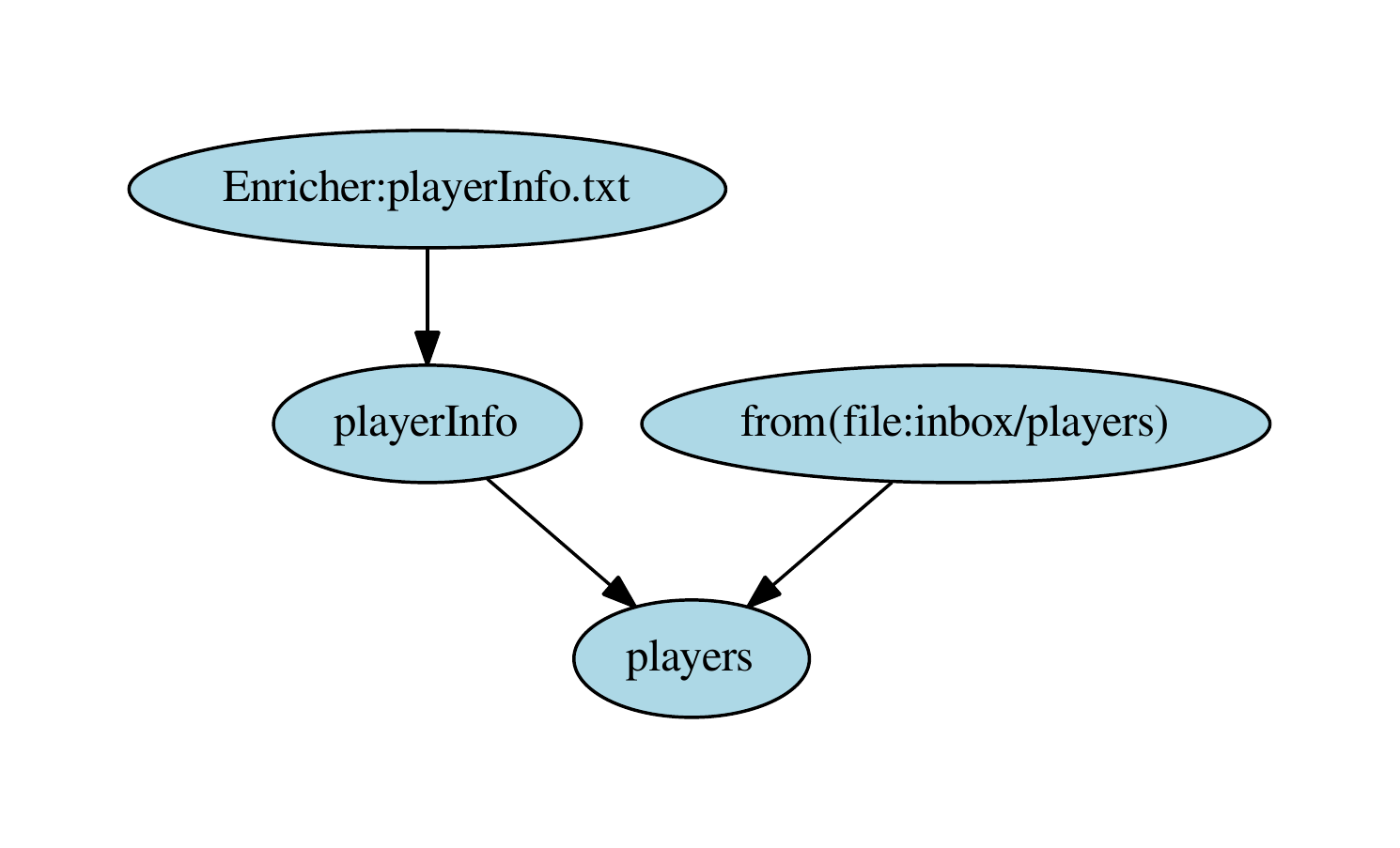}} 
	\subfloat[{Route Graph}\label{fig:enricher-detected}]{\includegraphics[width=0.5\textwidth]{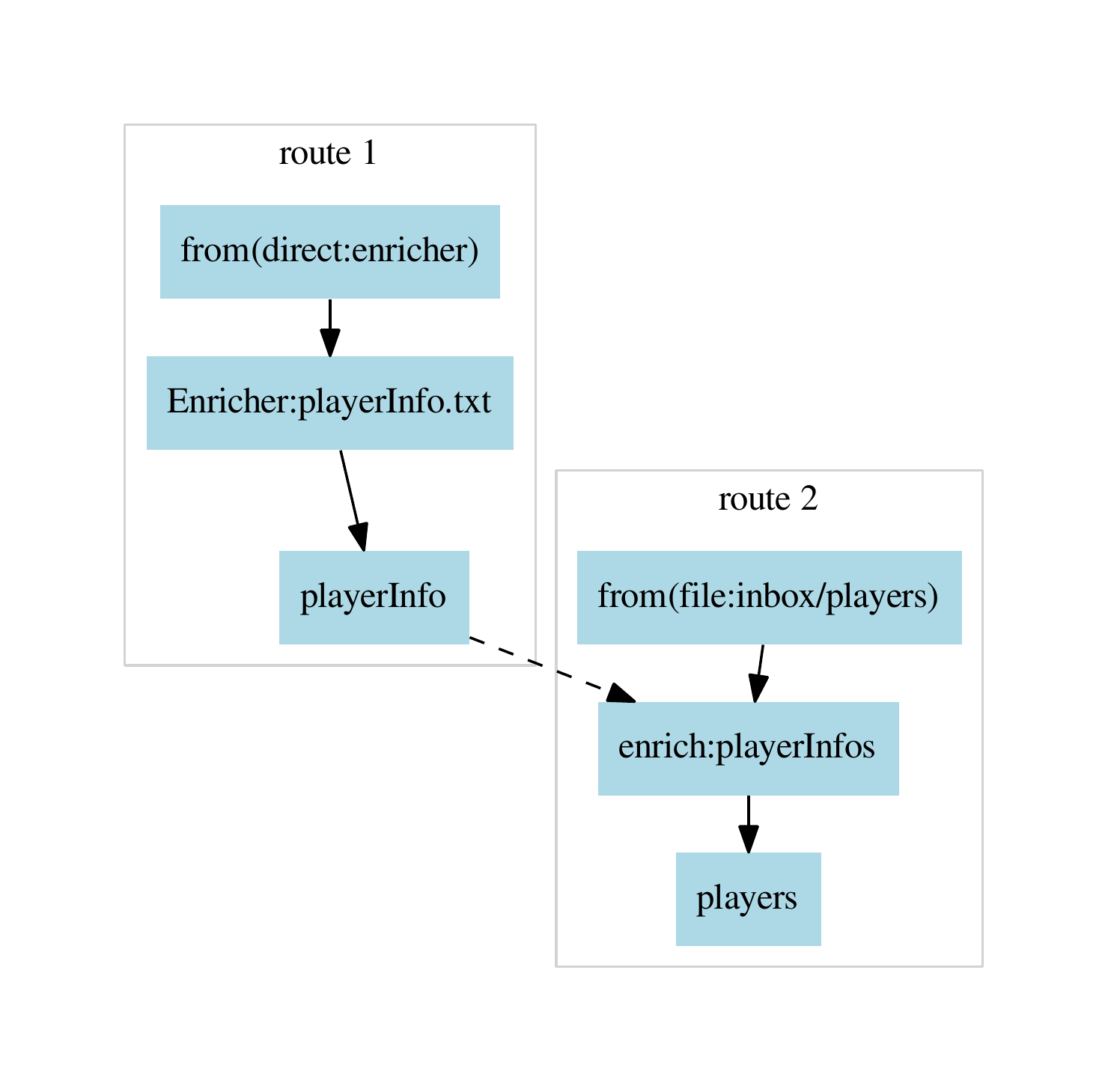}}
	\caption{Detection of the Remote Enricher Pattern}
	\label{fig:enricher-graph-rule}
\end{figure*}


\subsection{Messsage Channel Synthesis}
\begin{figure*}[h]
	\centering
	\includegraphics[width=0.9\textwidth]{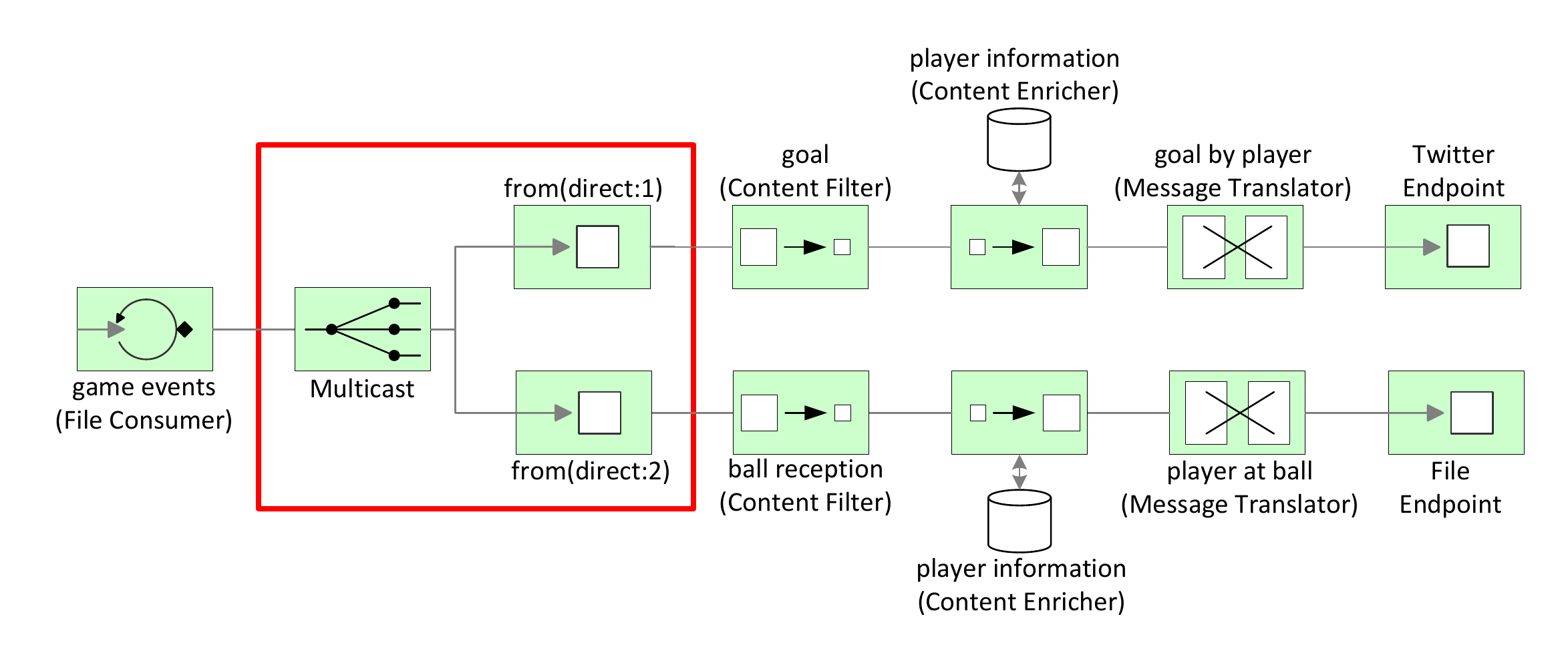}
	\caption{Generated Apache Camel routes in EIP-icon notation for the motivating example (inherent message filters before recipients omitted)}
	\label{fig:camel-routes-simple-prog}
\end{figure*}

The route graph represents the foundation for the code synthesis of the message channels, which are a combination of ILP constructs and Apache Camel patterns and routes. The construction of the routes is a trivial graph traversal starting from the fact source nodes. The multicast $t_{mu_1,2}$ and join router $t_{jr_1-3}$ transformations construct a $RG$ with $deg^-(n)==1$, with $n \in V_R$. Hence the ILP constructs can be synthesized one after the other based on their type and the ILP properties, which were preserved during the transformations and optimizations. For instance, Figure \ref{fig:camel-routes-simple-prog} shows the synthesized Apache Camel routes in the EIP icon notation. In comparison to the motivating example in Figure \ref{fig:twitter_bpmn}, the content-based router is exchanged by a multicast and message filters are added before the outbound message endpoints, while preserving the same semantics and allowing for parallel message processing.

\paragraph{Message Endpoints} The fact source and routing goal nodes are transformed to components in Apache Camel, passing the configurations that are stored in the node properties. A detected (not \texttt{@from} annotated) fact source gets an additional \texttt{numOfMsgsToAgg} property, which remembers the entering message count of a join router a corresponding aggregator ILP with \texttt{completionSize}=\texttt{numOfMsgsToAgg} is added. The \texttt{location} property defines the component's endpoint configuration and the \texttt{format} leads to the generation of a ILP format converter (\eg, JSON, CSV to Datalog) that is configured using the meta-facts supplied in the annotation body, conduction an additional projection. If the format is set to \texttt{datalog} no format conversion is needed.
The routing goals are configured similarly. A message filter ILP is added that discards empty messages. The a format converter (\eg, Datalog to JSON/CSV) is added and configured through the meta-facts property. Finally a Camel producer component is added to the route and configured.

\paragraph{Complex Routing Patterns} For the aggregator, additional \texttt{renamingRules} properties and renaming message translators are generated, containing a Datalog rule that adds \texttt{-aggregate} suffixes to every Datalog predicate used in the head of a query (for name differentiation).
Similarly, for the splitter, \texttt{-split} suffixes are generated that allow additional message translators to rename the predicates. This is necessary in order to build the dependency graph as described in section \ref{sec:lila}.

The inherent multicast nodes are configured through a recipient list property, containing the target node identifiers, which allows for a translation to the Camel multicast (no ILP defined).
\paragraph{Message Translation Patterns} The content filter and message translator nodes can be generated to the ILP content filter, which is based on a Camel processor and configured accordingly.
The node of the inherent content enricher, which can be specified by writing facts into a LiLa program, stores in the facts as properties. The
generated ILP (again based on a Camel processor) adds the facts to every incoming message. The explicit file enricher pattern is configured similar to a fact source, however, the configuration specifies a \texttt{fileName} property, used to configure the Camel component. Again, ILP format converters are added and configured by the meta-facts property.

\section{Experimental Evaluation} \label{sec:experimental}
We implemented ILP constructs as extensions to the lightweight, open source integration system Apache Camel in version 2.12.2 based on Section \ref{sec:datapatterns} that are references to LiLa programs as described in Section \ref{sec:synthesis}. The \emph{HLog} Datalog system we used for the measurements is a Java implementation of the standard na\"{i}ve-recursive Datalog evaluation (\ie, without stratification) from Ullman \cite{DBLP:books/cs/Ullman88} in version 0.0.6 as described in \cite{DBLP:conf/datalog/RitterW12}.

\subsection{Platform and Test Messages}
We conduct all measurements on a HP Z600 work station, equipped with two Intel Xeon processors clocked at 2.67GHz with a total of $2x6$ cores and $2x6$ logical processors, 24GB of main memory, running a 64-bit Suse Linux $11.4$ and a JDK version $7u71$.

In the experiments, messages with a single or multiple facts are used. The single fact JSON message is shown in Listing \ref{lst:single-fact-json} together with the corresponding Datalog program in Listing \ref{lst:single-fact-message}. The messages' payloads are approximately 20 and 41 bytes. Note that meta-facts are optional.

\begin{minipage}{0.5\textwidth}
	\begin{lstlisting}[language=LiLa,caption=Single-fact message,label={lst:single-fact-message}]
	match("true").
	meta("match","matching",1).
	\end{lstlisting}
\end{minipage}
\begin{minipage}{0.5\textwidth}
	\begin{lstlisting}[language=lila,caption=Single-entry message in JSON,label={lst:single-fact-json}]
	[{"matching" : "true"}]
	\end{lstlisting}
\end{minipage}


The multi-facts message tests were conducted with the JSON message shown in Listing \ref{lst:multi-fact-message-json} with its corresponding Datalog representation in Listing \ref{lst:multi-fact-message}. The messages' payloads are approximately $58$ and $85$ bytes, respectively.

\begin{minipage}{0.5\textwidth}
	\begin{lstlisting}[language=lila,caption=Mutli-fact message,label={lst:multi-fact-message}]
	match("true",1).
	match("false",2).
	meta("match","matching",1).
	meta("match","count",2).
	\end{lstlisting}
\end{minipage}
\begin{minipage}{0.5\textwidth}
	\begin{lstlisting}[language=lila,caption=Multi-entry message in JSON,label={lst:multi-fact-message-json}]
	[{"matching":"true", "count":1},
	{"matching":"false", "count":2}]
	\end{lstlisting}
\end{minipage}

\subsection{Performance of Logic Integration Patterns}
The subsequently described measurements target an experimental runtime evaluation of some of the introduced integration patterns from section \ref{sec:datapatterns} through a comparison of the ILP with the original Java-based implementation. Keep in mind, that tests with empty messages or routes without any processing steps result in identical performance results. All tests measure the pattern processing only, thus neglect the necessary format conversions.

\paragraph{Message Filter and Content-based Router Patterns} The basic router pattern analysis is conducted for the message filter and content-based router using the single-fact message for ILP (cf. Listing \ref{lst:single-fact-message}) and the corresponding JSON message for the Java implementation (cf. Listing \ref{lst:single-fact-json}). We execute the message filter in a Camel route for ILP (generated from a LiLa program) and for Java as shown in Listings \ref{lst:lila-message-filter-perf-route} and \ref{lst:lila-message-filter-perf-route-java}. The performance is measured without the message endpoints by sending multiple single fact messages, while one half of the messages is filtered out and the other half is routed further. 

\begin{lstlisting}[language=LiLa,caption=LiLa message filter performance test program,label={lst:lila-message-filter-perf-route}]
@from(file:data/testMessageFilter)
{match(matching).}
match-filtered(matching):-match("true").
@to(file:data/filtered)
\end{lstlisting}

\begin{lstlisting}[language=Lila,caption=Camel Route used for the message filter performance meassurements of Camel-Java,label={lst:lila-message-filter-perf-route-java}]
from(file:data/testMessageFilter)
.filter(new JsonMatchKeyValueExpression("match","true"))
.to(file:data/filtered);
\end{lstlisting}



\begin{figure}[h!]	
	\centering
	\includegraphics[width=0.45\textwidth]{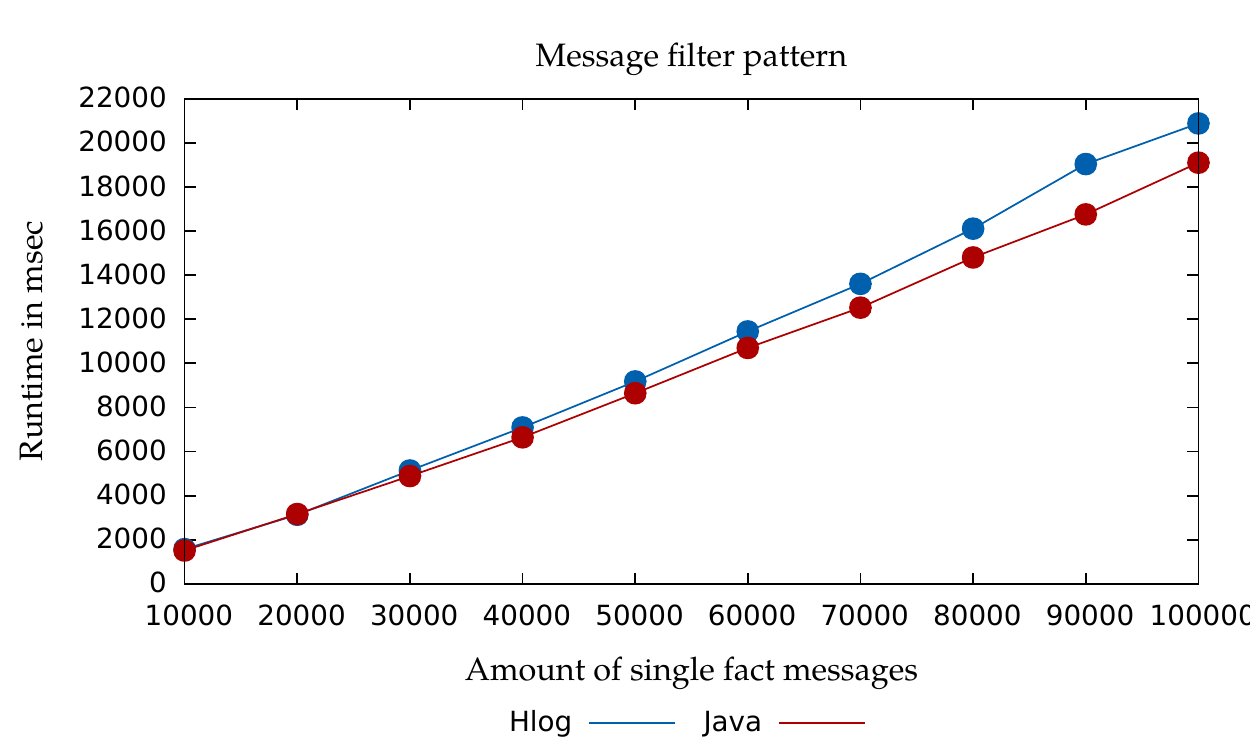}
	\hspace{0.05\textwidth}
	\caption[]{Basic Message Routing Pattern test}
	\label{fig:hlog-iris-content-based-router10k}
\end{figure}

The performance measurement results are depicted in Figure \ref{fig:hlog-iris-content-based-router10k}. The Camel-ILP and the Camel-Java implementation show a linear performance on the amount of incoming messages. Although the setup favors the Java processing due to (a) only single-fact messages are sent (\ie, Datalog evaluation is better suitable for set-operations), and (b) the type of operation during message routing is mostly only used to ``peak" into the message content (cf. Section \ref{sec:datapatterns}), the Java implementation seems to be only slightly better for higher amounts of single fact messages. A similar result/behavior can be observed for the content-based router pattern.

\paragraph{Content Filter Pattern} As an example for message transformations, we evaluate the content filter on a single message containing a varying amount of facts based on the message payload from Listings \ref{lst:multi-fact-message} and \ref{lst:multi-fact-message-json}. The routes in Listings \ref{lst:lila-content-filter-perf-route} and \ref{lst:lila-content-filter-perf-route-java} show that the content filter is configured with a single rule, for which half of the facts match.

\begin{lstlisting}[language=LiLa,caption=LiLa content filter performance test program,label={lst:lila-content-filter-perf-route}]
@from(file:data/testContentFilter)
{  match(matching,count). }

match-filtered(matching,count):-match("true",count).

@to(file:data/contentFilter)
{  match-filtered  }
\end{lstlisting}

\begin{figure}[h!]	
	\centering
	\hspace{0.05\textwidth}
	\includegraphics[width=0.45\textwidth]{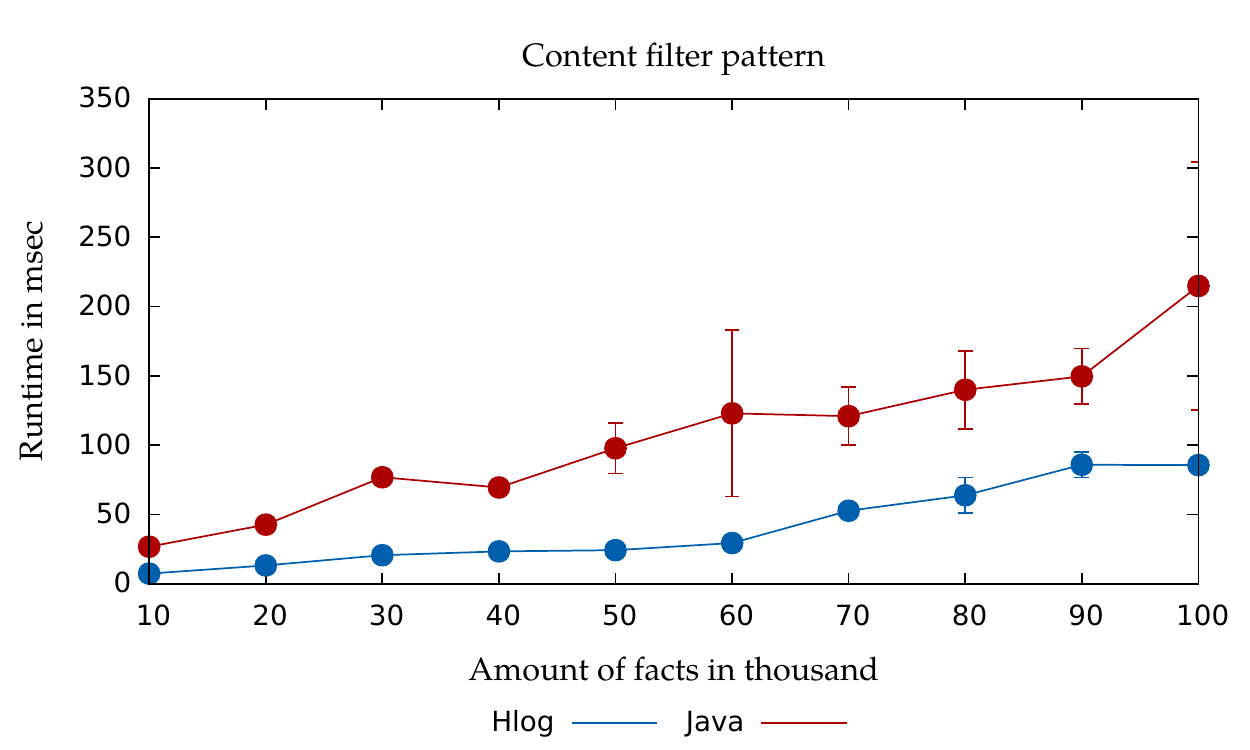}
	\caption[]{Basic Message transformation pattern test}
	\label{fig:aggregator-perf}
\end{figure}

\begin{lstlisting}[language=Lila,caption=Camel Route used for the content filter performance meassurements of Camel-Java,label={lst:lila-content-filter-perf-route-java}]
from(file:data/testContentFilter)
.process(new JSONContentFilter(new JsonMatchKeyValueExpression("match","true")))
.to(file:data/filtered);
\end{lstlisting}
The results of the measurement depicted in Figure \ref{fig:aggregator-perf} show a linear performance compared to the amount of facts processed. Noticeable, ILP is approximately twice as fast as the pure Camel-Java implementation. This is especially relevant for data-intensive processing scenarios and supports the observations (a,b) from the routing pattern measurement. Even if the multi-fact message contains only two facts, (a) the Datalog evaluation is already faster and (b) the approach favor more data-intensive operations on the message that are not only ``peaking" into the content for simple routing.

\subsection{``Soccer Player" Integration (revisited)}
Coming back to the motivating example in Listing \ref{lst:motivating-example}, which we extended with the calculation of the player's position (cf. \texttt{posAtShotOnGoal}), while shooting on goal, and to sample the player positions on a minute basis by using a recursive rule (cf. \texttt{pPosPerMinute}). Listing \ref{lst:extended-example} shows the extended version of the LiLa program that ``tweets" the calculated positions and stores stores them with the ``players at ball" to a file, and the positions per minute to a database. 

\begin{lstlisting}[language=lila,caption={Soccer Game Event Integration (revisited) as LiLa program.},label={lst:extended-example}]
@from(file:gameEvents.json,json)
{gE(period,time,eventCode,pId).}

@from(file:playerPosition.json,json)
{pPos(period,time,playerId,posX,posY).}

g(period,time,pId):-gE(period,time,"Goal",pId).
p(period,time,pId):-gE(period,time,"BallReception",pId).

gByP(period,time,pId,firstN,lastN):-g(period,time,pId),pInfo(pId,firstN,lastN).
pAtB(period,time,pId,firstN,lastN):-p(period,time,pId),pInfo(pId,firstN,lastN).

posAtShotOnGoal(period,time,firstN,lastN,posX,posY):-gByP(period,time,pId,firstN,lastN),pPos(period,time,pId,posX,posY).

pPosPerMinute(period,time,playerId,posX,posY):-pPos(period,millitime,posX,posY),time:=1,time=millitime/600.
pPosPerMinute(period,time,playerId,posX,posY):-pPos(period,millitime,posX,posY),pPosPerMinute(A,previousTime,B,C,D),time:=previousTime+1,time=millitime/600.

@enrich(playerInfo.json,json)
{pInfo(pId,firstN,last).}

@to(twitter:$config,json)
{gByP}

@to(file:playersAtBall.json)
{pAtB}

@to{file:positionAtShotOnGoal}
{posAtShotOnGoal}

@to{jdbc:soccerDatabase}
{pPosPerMinute}
\end{lstlisting} 

\begin{figure}[ht!]
	\centering
	\includegraphics[width=0.55\textwidth]{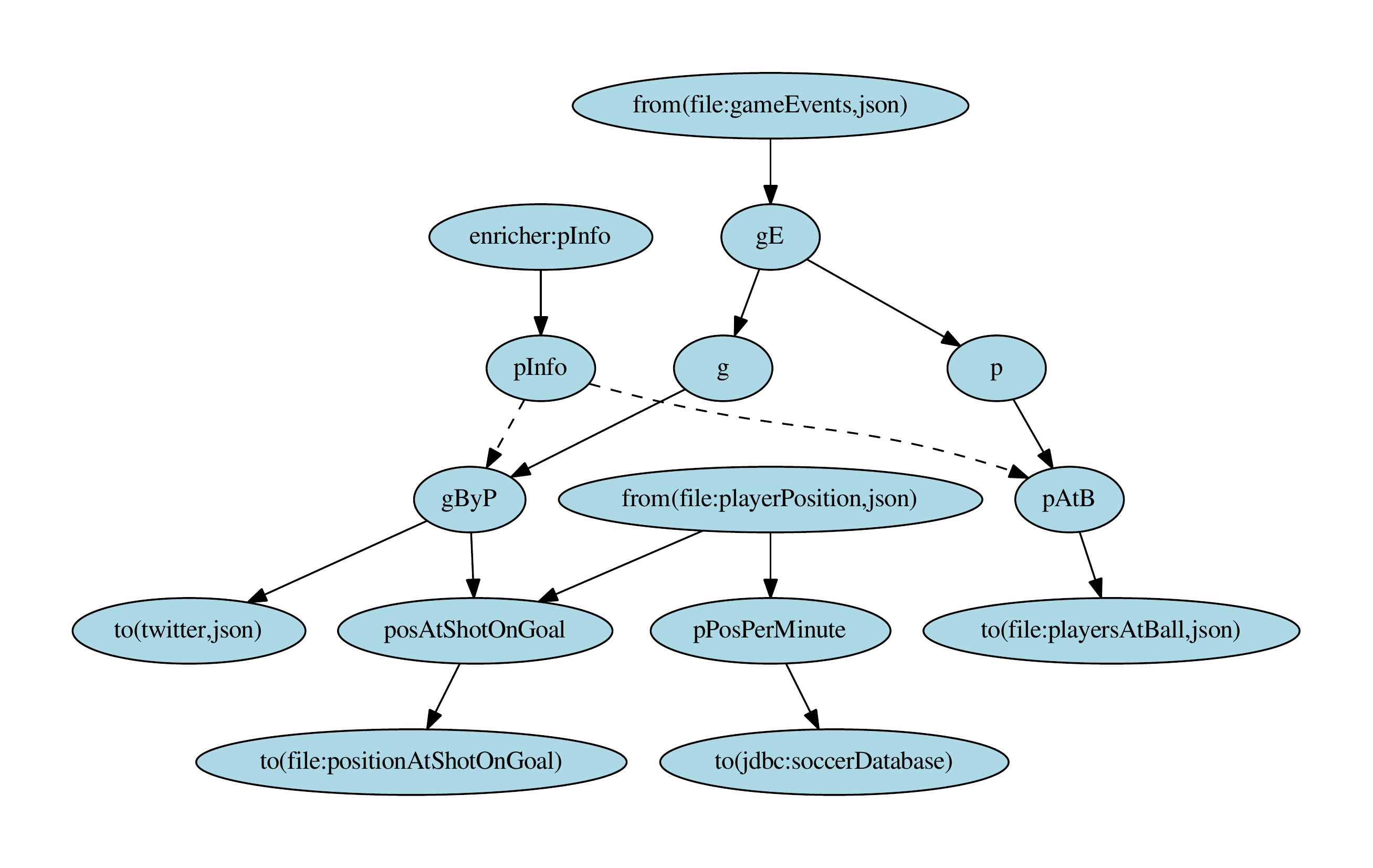}
	\caption[Data integration scenario dependency graph]{Dependency graph of the motivating example (revisited).}
	\label{fig:dep-graph-bundesliga}
\end{figure}

\begin{figure*}[ht!]
	\centering
	\includegraphics[width=0.9\textwidth]{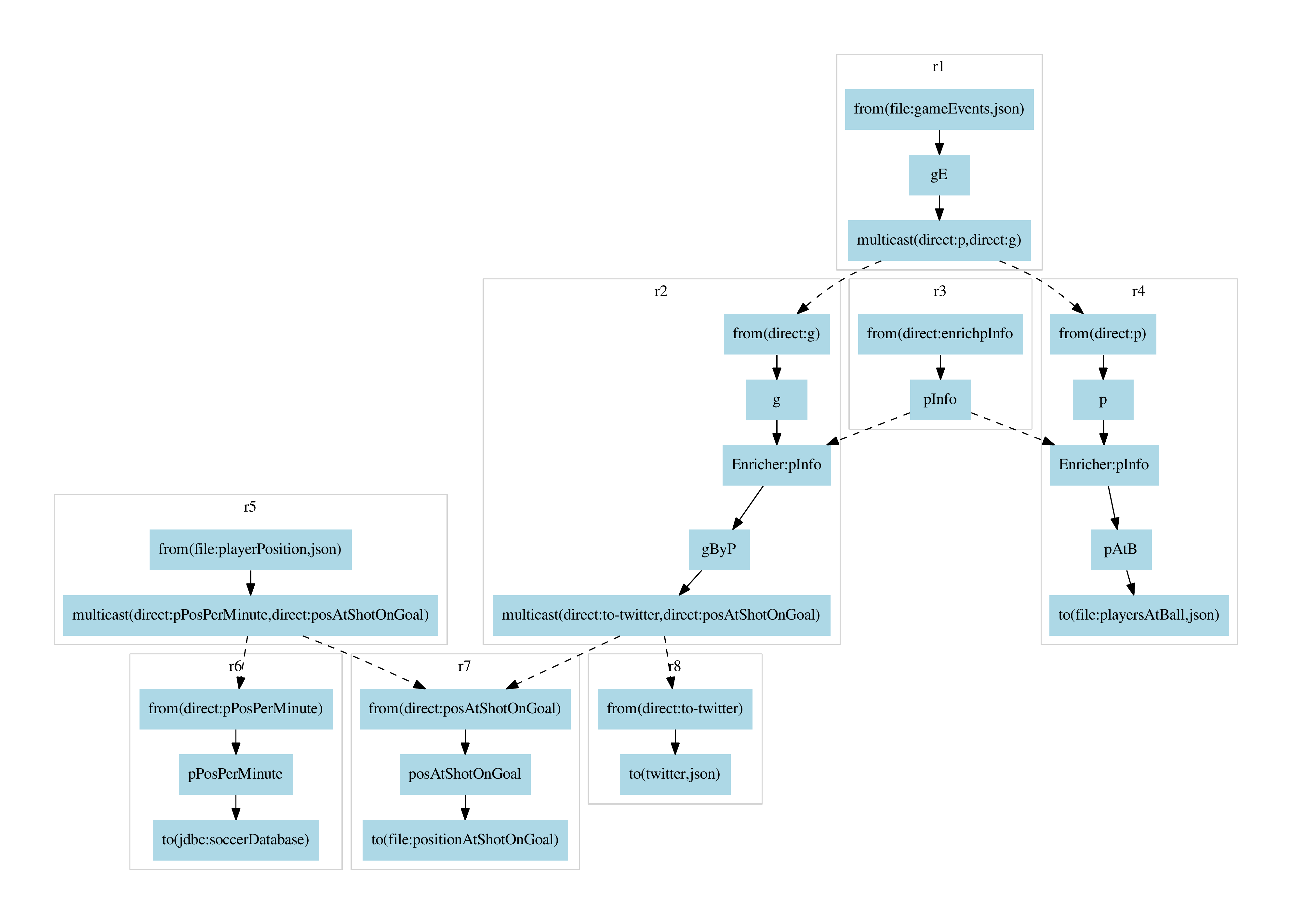}
	\caption[Data integration scenario route graph]{Route graph motivating example (revisited).}
	\label{fig:runtime-graph-bundesliga}
\end{figure*}

The corresponding (extended) LiLa dependency graph $LDG$ is shown in Figure \ref{fig:dep-graph-bundesliga}, which is used to generate the route graph $RG$ depicted in Figure \ref{fig:runtime-graph-bundesliga}. As the node \texttt{posAtShotOnGoal} has multiple incoming arcs, a join router pattern is detected and generated. Similarly a multicast pattern is detected and generated after the \texttt{from(file:playerPosition,json)} and \texttt{gByP} node.

\section{Related Work} \label{sec:relatedWork}
The application of Datalog to integration programming for current middleware systems has not been considered before and was only recently brought into talk by our BPMN-based modeling approach \cite{DBLP:conf/dexa/RitterB14}. However, the work on Java systems like Telegraph Dataflow \cite{DBLP:journals/sigmod/ShahMFH01}, Jaguar \cite{DBLP:journals/concurrency/WelshC00}) can be considered related work in the area of programming languages on application systems for faster, data-intensive processing. These approaches are mainly targeting to make Java better capable for data-intensive processing, while struggling with threading, garbage collection and memory management. None of them considers the combination of the host language with relational logic processing.

\paragraph{Declarative XML Processing and Semantic Web} Related work can be found in the area of declarative XML message processing (\eg, \cite{DBLP:conf/cidr/0002KM07}). Using an XQuery data store for defining persistent message queues, the work targets only a subset of ILP (\ie, persistent message queuing).

In the semantic web domain, several approaches use Datalog to integrate and query data from different mostly XML-based sources. For instance, the Semantic Web Integration Middleware (SWIM) extends Datalog with XPath expressions in the rule body to map XML to RDF as well as RQL to relational queries \cite{DBLP:conf/semweb/ChristophidesKKKMPST03}. ILP takes this approach one step further by using standard Datalog$^+$ to describe integration patterns that can be composed to integration scenarios.

\paragraph{Data Integration} The data integration domain uses integration middleware systems for querying remote data that is treated as local or ``virtual" relations. Starting with SQL-based approaches, \eg, using the \texttt{Garlic} integration system \cite{DBLP:conf/vldb/HaasKWY97}, the data integration research reached relational logic programming, summarized by \cite{DBLP:series/synthesis/2010Genesereth}. In contrast to remote queries in data integration, ILP extends integration programming with declarative, relational logic programming for application integration as well as the expressiveness of logic programs through integration semantics.



\paragraph{Declarative Application Programming}
With LiLa, we defined a language design similar to the trend of Datalog-style rule-based languages for declaratively, data-centric application development. Major work in this complementary field has been conducted by Green et al \cite{DBLP:conf/datalog/GreenAK12} with the $Datalog^{LB}$ language for (analytical) applications and Abiteboul et al \cite{DBLP:conf/sigmod/AbiteboulAMST13}, who applied logic programming (\ie, extended Datalog) to analytical and web application development and developed \emph{Webdamlog} \cite{DBLP:conf/pods/AbiteboulBGA11,DBLP:conf/sigmod/AbiteboulAMST13}, which is a language based on Datalog developed for specifying distributed applications. 

\paragraph{Data-aware Integration Languages}
The modeling of data-intensive workflows and integration scenarios has been approached only recently. Abiteboul et al compare business entity modeling to their Active XML (AXML) approach \cite{DBLP:conf/birthday/AbiteboulV13}. AXML is a data-aware workflow language, which specifies XML documents with embedded Web Service calls. Compared to LiLa, Active XML partly defines the notion of fact sources and content enrichers omitting message translation, routing goals and complex routing patterns like aggregator and splitter patterns. Another workflow approach is described in \cite{DBLP:conf/bis/PetruselVDM11}, which describes a decision mining approach that results to a \emph{Product Data Model} that strives to give insights into the data view of a business decision process. In LiLa the data graph is more explicit and the control flow model is of no concern to the user.

In the area of message-based integration, we define integration scenarios as BPMN-based \emph{Integration Flow} (IFlow), which specify the control-, data-, and exception-flow modeling \cite{DBLP:conf/ecmdafa/Ritter14,DBLP:journals/corr/Ritter14}. Although the IFlow approach is far better than control flow-centric models (\eg, Guaran\'{a} DSL \cite{DBLP:journals/ijcis/FrantzQC11}), data operations and formats still remain implicit.


\section{Conclusion and Future Work} \label{sec:conclusion}
According the observations \emph{P1--P4}, the main contributions of this work are (a) the analysis of the ``de-facto" standard integration patterns with respect to their enhancement for data-intensive processing, (b) the definition of integration logic programs, which are relational logic language constructs that can be embedded into patterns aligned with their semantics, (c) the definition of a data-aware logic integration language, which can be synthesized to integration logic programs, (d) an application to a conventional integration system, and (e) a brief performance analysis and the application to a data-intensive integration scenario.

Future work will be conducted in the area of rule-based optimization during the automatic program to runtime compilation for common integration processing styles, \eg, for scatter/gather, splitter/gather, with the related questions on data partitioning and provisioning during message processing.

\bibliographystyle{abbrv}
\bibliography{sigproc}


\balancecolumns

\end{document}